\documentclass[cmp]{svjourb}
\usepackage{pstricks}
\usepackage[utf8]{inputenc}
\usepackage{amsmath}
\usepackage{subfig}
\usepackage{amsmath, amssymb}
\usepackage{stmaryrd}
\usepackage{hyperref}

\journalname{Communications in Mathematical Physics}

\newcommand{\rD}{-4pt}

\newpsobject{showgrid}{psgrid}{subgriddiv=1,griddots=10,gridlabels=6pt}

\newcommand{\dVL}[1]{\raisebox{\rD}{\begin{pspicture}(-0.1,-0.3)(0.1,0.2)
\psline[linecolor=#1,linestyle=dotted](0,-0.3)(0,0.2)
\psdot[dotsize=1pt 2](0,-0.3)
\psdot[dotsize=1pt 2](0,0.2)
\end{pspicture}}}
\newcommand{\VL}[1]{\raisebox{\rD}{\begin{pspicture}(-0.1,-0.3)(0.1,0.2) 
\psline[linecolor=#1](0,-0.3)(0,0.2)
\psdot[dotsize=1pt 2](0,-0.3)
\psdot[dotsize=1pt 2](0,0.2)
\end{pspicture}}}
\newcommand{\DC}[1]{\raisebox{\rD}{\begin{pspicture}(-0.2,-0.3)(0.2,0.2) 
\psline[linecolor=#1,linestyle=dotted](-0.1,-0.3)(0.1,0.2)
\psline[linecolor=#1](0.1,-0.3)(-0.1,0.2)
\psdot[dotsize=1pt 2](0.1,-0.3)
\psdot[dotsize=1pt 2](0.1,0.2)
\psdot[dotsize=1pt 2](-0.1,-0.3)
\psdot[dotsize=1pt 2](-0.1,0.2)
\end{pspicture}}}
\newcommand{\DVL}{{\psset{doubleline=true}\VL{black}}}

\newcommand{\RpR}[2]{{\psframebox[fillcolor=white,fillstyle=solid]{
\(\;~+~\;\)
}
\psline[linecolor=#1](-1.2,0.2)(-0.9,0.2)
\psline[linecolor=#2](-1.2,0)(-0.9,0)
\psline(-1.05,-0.13)(-1.05,0.31)
\psdot(-1.05,0.2)
\rput(0.7,0){\psline[linecolor=#1](-1.2,0.2)(-0.9,0.2)
\psline[linecolor=#2](-1.2,0)(-0.9,0)
\psline(-1.05,-0.13)(-1.05,0.31)
\psdot(-1.05,0)
}
}}

\newcommand{\x}{{    x}}
\newcommand{\z}{{    z}}
\newcommand{\PPi}{{    \Pi}}
\renewcommand{\t}{{    t}}

\newcommand{\diag}{\mathrm{diag}}

\newcommand{\I}{\mathbb{I}}
\newcommand{\F}{\mathfrak{I}}
\newcommand{\eG}[1]{#1}
\renewcommand{\P}{\mathcal{P}}
\newcommand{\proj}{P}
\newcommand{\bI}{{\overline{I}}}

\def\l{\lambda}
\def\a{\alpha}
\def\b{\beta}
 \def\th{\theta}
 \def\de{\delta}

\def\<{\langle}
\def\>{\rangle}

\def\sdet{{\rm sdet}~}

\def\hD{\hat D}

\def\tr{{\rm tr~}}

\newcommand{\W}{{\mathbf W}}
\newcommand{\Qb}{{\mathbf Q}}
\newcommand{\Qs}{{\mathsf Q}}
\newcommand{\Tb}{{\mathbf T}}

\newcommand{\thG}{GL(K)}

\newcommand{\CF}{w(\x)}

\newcommand{\gtD}[1]{{\raisebox{-7pt}{
\renewcommand{\ll}{0}\newcommand{\yo}{0.2}\newcommand{\rr}{2}   \newcommand{\yt}{0.5}\newcommand{\dd}{0}\newcommand{\xo}{0.3}\newcommand{\uu}{0.7} \newcommand{\xt}{.8}\newcommand{\ro}{1.2}\begin{pspicture}(\ll,\dd)(\rr,\uu)
          \psline[linecolor=blue](\ll,\yo)(\ro,\yo)
          \psline[linecolor=red](\ll,\yt)(\ro,\yt)
          \psline(\xo,\dd)(\xo,\uu)
          \psline(\xt,\dd)(\xt,\uu)
          \rput(1.7,\yt){\(_{\PPi}\)}
          \rput(1.7,\yo){\(_{\CF}\)}
          #1
        \end{pspicture}}}~~}

\begin{document}


\title{ Baxter's Q-operators and operatorial B\"acklund flow for quantum
  (super)-spin chains }

\author{Vladimir Kazakov\inst{1}\fnmsep\inst{2}\fnmsep\thanks{member of
    Institut Universitaire de France}, Sebastien
  Leurent\inst{1}, and Zengo Tsuboi\inst{3}\fnmsep\inst{4}\fnmsep\thanks{Present address:
 Institut f\"{u}r Mathematik und Institut f\"{u}r Physik, 
Humboldt-Universit\"{a}t zu Berlin, 
Johann von Neumann-Haus, 
Rudower Chaussee 25, 12489 Berlin, Germany}}

\institute{Ecole Normale Superieure, LPT,  75231 Paris CEDEX-5, France
  \and
Universit\'e Paris-VI, Paris, France
\and
Max-Planck-Institut f\"{u}r Gravitationsphysik, 
Albert-Einstein-Institut, 
Am M\"{u}hlenberg 1, 
14476 Potsdam,  
Germany
\and
Osaka City University Advanced Mathematical Institute,
3-3-138 Sugimoto, Sumiyoshi-ku
Osaka 558-8585 Japan
}

\psset{dotsep=1.3pt}

\titlerunning{Baxter's Q-operators and operatorial B\"acklund flow}

\maketitle

\begin{abstract}
We propose the operatorial Baxter's TQ-relations in a general
form of the operatorial B\"acklund flow describing the nesting process
for the inhomogeneous rational \(gl(K|M)\) quantum (super)spin chains
with twisted periodic boundary conditions. The full set of Q-operators
and T-operators on all levels of nesting is explicitly defined. The
results are based on  a generalization of the identities  among the
group characters and their group co-derivatives with respect to the
twist matrix, found by one of the authors and
P.Vieira\cite{Kazakov:2007na}.
Our formalism, based on this new ``master''
identity, allows a systematic and rather
straightforward derivation of the whole set of nested Bethe ansatz
equations for the spectrum of  quantum integrable spin chains,
starting from the R-matrix.   
\end{abstract}

 \pagebreak{}

\section{Introduction}

It has been noticed long ago that the  mathematical structures behind
the quantum  integrable spin chains have many  similarities with the
theory of  classical integrable systems, such as KP or KdV
hierarchies. It goes of course not only about the obvious
correspondence between the   quantum integrable systems and their
classical limits when, for example, the quantum transfer matrix
 of a
quantum 1+1-dimensional system becomes the classical  monodromy matrix
of the corresponding classical Lax connection. There is a more
striking ``classical'' feature of the quantum integrability: The
quantum transfer matrix represents a natural (spectral parameter
dependent) generalization of the Schur character of  a classical
algebra \cite{Bazhanov:1989yk} given by the so called
Bazhanov-Reshetikhin (BR) determinant formula and, as such, it
satisfies a certain Hirota bilinear finite difference equation, which
appears in the quantum context as a certain fusion relation among the
composite quantum states appearing in  quantum spin chains as certain
bound states (``strings'')  of Bethe roots
\cite{Pearce:1991ty,Kuniba:1993cn,Kuniba:1995vc,Bazhanov:1996dr,Tsuboi:1997iq}. Similar,
though a more complicated realization of Hirota discrete ``classical''
integrable dynamics has been observed in the context of the quantum
(1+1)-dimensional QFT's, or sigma-models
\cite{Bazhanov:1996dr,Gromov:2008gj,Kazakov:2010kf}, an 
observation which appeared to be at the heart of an important advance in the
study of the spectrum of the AdS\(_5\)/CFT\(_4\) correspondence
\cite{Gromov:2009tv,Bombardelli:2009ns}.  

Hirota equation immediately brings us to the idea that quantum integrability, at least for certain quantities, such as transfer-matrix eigenvalues, can be viewed as a specific case of classical integrability and of the theory of classical tau-functions. Indeed, the character of a classical group, say   \(gl(K)\), is nothing but a tau-function of the KdV hierarchy. It was proposed in \cite{Kazakov:2007na} to view the quantum transfer matrix of a rational quantum Heisenberg-type \(gl(K|M)\) (super)-spin chain with twisted boundary conditions as a quantum, operatorial generalization of the character and to construct the transfer matrix (T-operator)  by acting on the character in a given irrep, as a function of the  group element (twist), by  special  group derivatives, called the co-derivatives. The formalism of co-derivatives has led  to a direct proof of the BR formula \cite{Kazakov:2007na}  (see also \cite{Cherednik}), and the basic underlying identity for the characters found in \cite{Kazakov:2007na} seems to be just a new form of the KdV Hirota identity (the fact yet to be understood). 

In the present paper, we want to move even further in this classical
interpretation of the quantum integrability and to generalize the
basic identity of \cite{Kazakov:2007na} to include the Baxter's TQ-relations into our
formalism. This implies a natural definition of all Baxter's
Q-operators, rather different from the one 
known in the literature \cite{Bax72,Bazhanov:1996dr,Bazhanov:1998dq,BaxQ-papers,%
Bazhanov:2001xm,Belitsky:2006cp,Bazhanov:2008yc,BFLMS10}, 
and more generally, of the T-operators on all levels of the nesting procedure. This nesting takes
a form of a B\"acklund flow, directly for the T- and Q-operators. Due
to the fact that all of them belong to a commuting family of operators, all these relations can be immediately transformed into the well known functional form, for their eigenvalues 
\cite{Kazakov:2007fy,Krichever:1996qd,Zabrodin:2007rq,Tsuboi:2009ud}.

Our main identity given in the next section offers an interesting alternative  and a concise approach  to the quantum integrability  uncovering the whole structure of the nested Bethe ansatz,  from the R-matrix and the Yang-Baxter relations all the way to the nested Bethe ansatz equations, in the  general operatorial form for all the intermediate quantities.

\section{Transfer-matrix, co-derivative and TQ-relations}

 We recall that the main object of our study is the transfer matrix of an inhomogeneous quantum spin chain\footnote{
  Throughout this paper, all operators will act on the same
  Hilbert 
  ``quantum'' space \(\mathcal{H}=(\mathbb{C}^K)^{\otimes N}\) (resp
  \((\mathbb{C}^{K|M})^{\otimes N}\) for supersymmetric spin
  chains).
We use the Young diagram 
 \(\l\) to label the irreducible tensor representation. 
The fundamental representation corresponds to 
the Young diagram with one box.
}  
\begin{equation}\label{TMATR}
    \Tb^{\{\lambda\}}(u)={\rm tr}_\lambda  \left({\mathbf R}_{N}^{\{\lambda\}}(u-\theta_N)\otimes \dots
    \otimes {\mathbf R}_{1}^{\{\lambda\}}(u-\theta_1)\, \pi_\lambda(g) \right)
\end{equation}
where \(\, \pi_\lambda(g)\) is a matrix element of  a twist matrix
\(g\in \thG\)
in an irrep
 \(\lambda\) and
 \begin{equation}
    {\mathbf R}_i^{\{\lambda\}} (u)= u~\I
      +2\sum_{\a\b}  e_{\b\a}^{(i)}  \otimes
    \pi_\l(e_{\a\b})\equiv u+2~\P_{i,\l}  \label{RMATR}
\end{equation}
is the R-matrix in irrep 
 \(\l\) in auxiliary space (and in fundamental
irrep in the quantum space). The identity operator \(\I\) is implicit
in the r.h.s., it will
  often be  omitted for the brevity. The \(gl(K)\)  generator \( e_{\a\b}^{(i)}\)   corresponds to the
\(i^{\mathrm{th}}\) quantum space\footnote{Here 
  \(\{e_{ij}\}_{i,j=1}^{K}\) satisfy the commutation relations  
\([e_{ij},e_{kl}]=\delta_{jk}e_{il}-\delta_{li}e_{kj}\).
  The
    \(i^{\mathrm{th}}\) quantum space is the \(i^{\mathrm{th}}\)
    factor in 
    \(\mathcal{H}=\underbrace{(\mathbb{C}^K)\otimes(\mathbb{C}^K)
    \otimes\cdots\otimes(\mathbb{C}^K)}_{\textrm{\(N\)
      times}}\). In this way, e.g.  
\(e_{\b\a}^{(i)} =\I^{\otimes (i-1)} \otimes e_{\b\a} \otimes \I^{\otimes (N-i)} \).
The generalization of our
  construction to the case of \(gl(K|M) \) super-spins will be given
  in sec.\ref{sec:super}.} (which is in the fundamental
representation) and  \(\pi_\l(e_{\a\b})\) - to the
auxiliary space. When \(\lambda \) is also fundamental then 
   \(\P_{i,\l}\) becomes a usual permutation operator 
\(e_{\b\a}^{(i)} =\I^{\otimes (i-1)} \otimes e_{\b\a} \otimes \I^{\otimes (N-i)}\)
so that
  \(\P_i=\sum_{\a\b} e_{\b\a}^{(i)}  \otimes  e_{\a\b}\). 
   \(\P_{i}\) 
permutes
 the indices of the auxiliary space and the
   quantum subspace. 
This \(\Tb^{\{\lambda\}}(u) \) is a spectral parameter \(u \in {\mathbb C}\) 
dependent operator on the quantum
space \(\mathcal{H}\). It is polynomial in \(u\) and in the
inhomogeneities \(\th_i \in {\mathbb C} \).

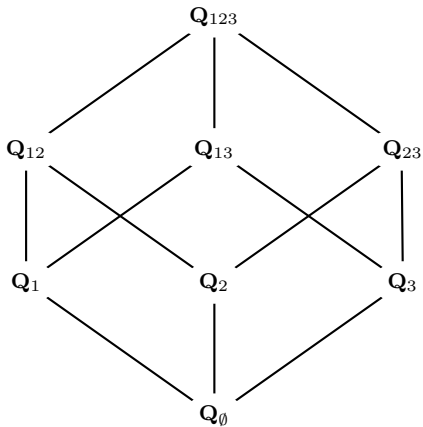
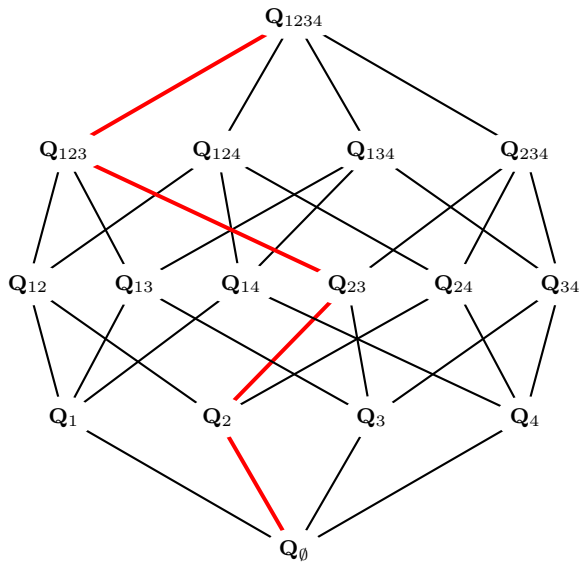
\begin{figure}
  \centering
\psset{unit=1pt}
\subfloat[Hasse diagram for the Q-operators: \(gl(3)\) case. 
 There are \(2^3=8\) Q-operators.]{\label{Hassegl3}
\begin{pspicture}(-10,-10)(160,160)
\psline(75,0)(4.3,50)
\psline(75,0)(75,50)
\psline(75,0)(145.8,50)
\psline(4.3,50)(4.3,100)
\psline(4.3,50)(75,100)
\psline(75,50)(4.3,100)
\psline(75,50)(145.8,100)
\psline(145.8,50)(75,100)
\psline(145.8,50)(145.3,100)
\psline(4.3,100)(75,150)
\psline(75,100)(75,150)
\psline(145.8,100)(75,150)
\rput[c]{0}(75,0){\psovalbox*{\(\Qb_{\emptyset}\)}}
\rput[c]{0}(4.3,50){\psovalbox*{\(\Qb_{1}\)}}
\rput[c]{0}(75,50){\psovalbox*{\(\Qb_{2}\)}}
\rput[c]{0}(145.8,50){\psovalbox*{\(\Qb_{3}\)}}
\rput[c]{0}(4.3,100){\psovalbox*{\(\Qb_{12}\)}}
\rput[c]{0}(75,100){\psovalbox*{\(\Qb_{13}\)}}
\rput[c]{0}(145.8,100){\psovalbox*{\(\Qb_{23}\)}}

\rput[c]{0}(75,150){\psovalbox*{\(\Qb_{123}\)}}

\end{pspicture}}
\qquad
\subfloat[Hasse diagram for the Q-operators: \(gl(4)\) case. 
 There are \(2^4=16\) Q-operators. ]{\label{Hassegl4}
\begin{pspicture}(-10,-10)(210,210)
\psline(100,0)(13.4,50)
\psline[linecolor=red,linewidth=1.5](100,0)(71.1,50)
\psline(100,0)(128.9,50)
\psline(100,0)(186.6,50)
\psline(13.4,50)(0,100)
\psline(13.4,50)(40,100)
\psline(13.4,50)(80,100)
\psline(71.1,50)(0,100)
\psline[linecolor=red,linewidth=1.5](71.1,50)(120,100)
\psline(71.1,50)(160,100)
\psline(128.9,50)(40,100)
\psline(128.9,50)(120,100)
\psline(128.9,50)(200,100)
\psline(186.6,50)(80,100)
\psline(186.6,50)(160,100)
\psline(186.6,50)(200,100)
\psline(0,100)(13.4,150)
\psline(0,100)(71.1,150)
\psline(40,100)(13.4,150)
\psline(40,100)(128.9,150)
\psline(80,100)(71.1,150)
\psline(80,100)(128.9,150)
\psline[linecolor=red,linewidth=1.5](120,100)(13.4,150)
\psline(120,100)(186.6,150)
\psline(160,100)(71.1,150)
\psline(160,100)(186.6,150)
\psline(200,100)(128.9,150)
\psline(200,100)(186.6,150)
\psline[linecolor=red,linewidth=1.5](13.4,150)(100,200)
\psline(71.1,150)(100,200)
\psline(128.9,150)(100,200)
\psline(186.6,150)(100,200)
\rput[c]{0}(100,0){\psovalbox*{\(\Qb_{\emptyset}\)}}
\rput[c]{0}(13.4,50){\psovalbox*{\(\Qb_{1}\)}}
\rput[c]{0}(71.1,50){\psovalbox*{\(\Qb_{2}\)}}
\rput[c]{0}(128.9,50){\psovalbox*{\(\Qb_{3}\)}}
\rput[c]{0}(186.6,50){\psovalbox*{\(\Qb_{4}\)}}
\rput[c]{0}(0,100){\psovalbox*{\(\Qb_{12}\)}}
\rput[c]{0}(40,100){\psovalbox*{\(\Qb_{13}\)}}
\rput[c]{0}(80,100){\psovalbox*{\(\Qb_{14}\)}}
\rput[c]{0}(120,100){\psovalbox*{\(\Qb_{23}\)}}
\rput[c]{0}(160,100){\psovalbox*{\(\Qb_{24}\)}}
\rput[c]{0}(200,100){\psovalbox*{\(\Qb_{34}\)}}
\rput[c]{0}(13.4,150){\psovalbox*{\(\Qb_{123}\)}}
\rput[c]{0}(71.1,150){\psovalbox*{\(\Qb_{124}\)}}
\rput[c]{0}(128.9,150){\psovalbox*{\(\Qb_{134}\)}}
\rput[c]{0}(186.6,150){\psovalbox*{\(\Qb_{234}\)}}
\rput[c]{0}(100,200){\psovalbox*{\(\Qb_{1234}\)}}
\end{pspicture}
}

 \caption{Hasse diagram for the Q-operators in \(gl(3)\) and gl(4)
   case. 
In (b), the red thick lines denote one of the \(4!=24\) nesting paths characterized by 
 a chain of the index sets \(\F=I_{4} \supset I_{3} \supset I_{2} \supset I_{1} \supset I_{0}\), 
 where 
\(I_{4}=\{1,2,3,4\}, I_{3}=\{1,2,3\}, I_{2}=\{2,3\}, I_{1}=\{2\},I_{0}=\emptyset\). 
Note that (b) contains the diagram (a) for \(gl(3)\) as a subdiagram.
}
  \label{Hassegl34}
\end{figure}

The main goal is to find all the eigenvalues of this transfer matrix for the \(gl(K)\) quantum spins.
For that we work out  an operatorial B\"acklund transformation, which can be  also called  the nesting procedure, whose main goal is to derive, in a deductive way, without any assumptions, the nested system of Bethe ansatz equations defining these eigenvalues. On the way, we will encounter a collection of the intermediate T-operators, and the Baxter's Q-operators as their particular case, at each level of nesting. The operatorial TQ-relations, representing the B\"acklund transformation reducing the problem  for a \(gl(k)\)  subalgebra to a similar problem for the \(gl(k-1)\) subalgebra, in the nesting procedure corresponding to the chain of embeddings \(gl(K)\supset gl(K-1)\supset\cdots\supset gl(1)\) were given in their functional form in  \cite{Krichever:1996qd} (for the super-spin chains in  \cite{Kazakov:2007fy}). 
All the T- and Q- operators in this nesting procedure are labelled by the subsets \(I\) of the 
full set \(\F =\{1,2,\dots, K \} \) as \(\Tb^{\{\lambda\}}_{I}(u)\) and \(\Qb_{I} (u) \).  
The original transfer matrix \eqref{RMATR} corresponds to the T-operator for the full set 
\(\Tb^{\{\lambda\}}(u)=\Tb^{\{\lambda\}}_{\F}(u) \).  
There are \(2^{K}\) subsets of the full set \(\F \) and 
they can be described in terms of the Hasse diagram based on the
inclusion relations.  
A chain of subsets of the full set 
\(\F= I_{K} \supset I_{K-1} \supset \cdots \supset I_{0}=\emptyset \), where  
\(|I_k|\equiv\mathrm{Card}(I_{k})=k\), forms a path on the 
Hasse diagram. We will call this the {\it nesting path}, and 
\((K-k)\) - the {\it level} of  nesting. There are \(K!\) different nesting paths for \(gl(K)\) 
(see fig \ref{Hassegl4}).  
This description of all the \(2^{K}\) Q-operators based on the Hasse diagram 
was proposed in \cite{Tsuboi:2009ud}, and will be used throughout this paper. 

The  general TQ-relations,  derived in sec.\ref{sec:TQ} in the operatorial form, are given by

\begin{equation}
  \label{eq:TQfun}
  \Tb_{I}^{s}(u)\Qb_{I,j} (u)=
  \Tb_{I,j}^{s}(u)\Qb_{I}(u)-x_j
  \Tb_{I,j}^{s-1}(u+2)\Qb_{I}(u-2) ,
\end{equation}
where \(g=\diag(x_1,x_2,\cdots,x_K)\) is the twist matrix in the diagonal basis,
the superscript \(s\) in T-operator denotes the
symmetric \(\lambda= (s)\) irrep in the auxiliary space, by
\(I\subset\F=\{1,2,\cdots,K\}\) we denote  a subset of the full set of
indices (labeling  the eigenvalues) and by \(I,j\equiv I\cup\{j\}\subset \F\) we denote a subset with one
  more index \(j\notin I\).
This TQ-relation 
relates the T-operator \(\Tb_I^s\) and the T-operator
\(\Tb_{{I,j}}^{s}\) of the previous level of nesting (which has one more
index).
 A chain of these relations
allows to relate the original transfer matrix
\(\Tb_{\F}^{s}(u)\) on 
the level zero 
of nesting to
the \(u\)-independent operator
\(\Tb_{\emptyset}^{s}(u) \) given by  \eqref{eq:TQempty}. We will also show that the
  Q-operators are 
equal to the T-operators taken at an empty Young diagram:  
\begin{equation}\label{eq:QthroughT}
\Qb_{I}(u)=\Tb_{I}^0(u). 
\end{equation}

In the papers  \cite{Krichever:1996qd,Kazakov:2007fy}, all the   T- and Q-operators at intermediate steps were assumed, by self-consistency and without a proof, to be polynomials in \(u\). This analyticity assumption immediately leads to the nested Bethe ansatz equations defined by the nesting path. In this paper,   we complete the missing link of the chain and find  the explicit operatorial form of the B\"acklund flow \eqref{eq:TQfun}.

In what follows, we will extensively use the definitions and the
identities of \cite{Kazakov:2007na}. In particular, the  co-derivative \(\hD\) defined there and   used through the whole current paper is a very simple object defined by its action   on any function of \(g\) as follows
\begin{equation}
 \left.   \hD \otimes f(g)= \frac{\partial}{\partial\phi} \otimes
 f(e^{\phi\cdot e}g)\right|_{\phi=0} \label{Ddef}
\end{equation}
where   \(\phi\) is a matrix in the fundamental representation: \(\phi\cdot
e\equiv \sum_{\alpha\beta} e_{\alpha\beta} \phi^{\alpha}_{\,\beta}\) 
and \(\frac{\partial}{\partial\phi} =\sum_{\alpha \beta}e_{\alpha \beta} 
\frac{\partial}{\partial\phi^{\beta}_{\alpha}} \). 
Its main property, which also could serve as its definition, manifests in its action on the group element in fundamental irrep:
\begin{equation}
 \hD \otimes
g= \mathcal{P} \,(1\otimes g) \,\nonumber
\end{equation}
where \(\P\) is the operator of permutation between the 1\(^{\rm st}\) and the
 2\(^{\rm nd}\)  spaces\footnote{Explicitly in indices, the last relation looks like \(\hD^{i_1}_{j_1}\,\, g^{i_2}_{j_2}=\delta^{i_2}_{j_1}\,\,g^{i_1}_{j_2}\). 
 It is a usual matrix derivative obeying the Leibniz rules. 
Throughout this paper, we define the tensor (or matrix) indices of any operator \({\bf A}\) 
with respect to a basis \(\{ v_{k_1}\otimes v_{k_2} \otimes \cdots \otimes v_{k_N} \} \) as 
\({\bf A}v_{l_1}\otimes v_{l_2} \otimes \cdots \otimes v_{l_N} =
\sum_{k_1,k_2,\cdots ,k_N}
A_{l_1,l_2,\cdots ,l_N}^{k_1,k_2,\cdots ,k_N}  v_{k_1}\otimes v_{k_2} \otimes \cdots \otimes v_{k_N} \). 
In particular, a usual notation of a matrix element \(A_{k,l} \) is written as 
\(A^{k}_{l} \). 
}.

Many other useful  properties of this co-derivative, mostly following
from the application of the standard Leibniz rule can be found
in \cite{Kazakov:2007na}  and some of them are summarized in the appendix
\ref{sec:diagrammatics}.

Using the co-derivative we can for example  rewrite the T-operator \eqref{TMATR} in the following way:
\begin{equation}\label{DIFFT}
\Tb^{\{\l\}}(u) = (u_1+2 \hat D)\otimes(u_2+2 \hat D)\otimes\dots
  \otimes(u_N+2 \hat
   D)\,\,\,\chi_{_{\{\l\}}}(g)
\end{equation}
where \(u_i=u-\th_i\) and  \(\chi_{_{\{\l\}}}(g)=\tr\pi_\l(g)\) is the character of the twist
\(g\) in the irrep \(\l\).
The action of each of  \(N\) brackets adds a new spin to the
system, with its fundamental quantum space. 

\section{The master identity}

In this section we will formulate the  main identity of this
paper, called here the master identity \eqref{eq:TQNu} - the basis of our
approach to the quantum integrability.
It is proven in the appendix \ref{sec:proof-mast-ident}.

This identity involves the generating function of characters in  symmetric (\(\l=(s)\)) or antisymmetric
(\(\l=(1^{a})\)) tensor irreps,
\begin{equation}\label{eq:GenCh}
w(z)=\det\frac{1}{1-z~g}=\frac{1}{\prod_{j=1}^K(1-z~x_j)}
=\sum_{s=0}^\infty z^s~\chi_s(g)=\frac{1}{\sum_{a=0}^\infty (-1)^a z^a~\chi^{(a)}(g)},
\end{equation}
where \(x_j\) are the eigenvalues\footnote{Throughout
  this paper, we shall assume that \(\forall i\neq j\), \(x_i\neq
  x_j\).} of the twist matrix \(g\), 
and \(z \in {\mathbb C}\).
This master identity will relate  operators of the form
\(\otimes_i(u_i+2\hD)~\prod_k w(t_k)\) (for an arbitrary set of complex
numbers \( \{t_k \} \)),
 which is a generalization\footnote{The T-operator  in symmetric
   irrep can be indeed obtained as
\(\Tb^{s}(u) = \frac 1 {s!}\left(\frac{\partial}{\partial
      z}\right)^s 
\left[\otimes_i(u_i+2\hD)~w(z)\right]_{z\to 0}\).
} of the T-operator
\eqref{DIFFT}. To avoid the bulky notations in this definition, we
assume all the terms like \(u_i\) and \(2+u_i\)  
to be multiplied by the identity operator \(\mathbb{I}\), and the tensor product \(\otimes_i
\) is taken as\footnote{ \(\overrightarrow{\otimes}_{i=1}^{N}A_{i}=A_{1} \otimes A_{2} \otimes \cdots \otimes A_{N} \) 
for any indexed operators \(\{ A_{i} \}_{i=1}^{N}\).} 
 \(\overrightarrow{\otimes}_{i=1}^{N}\) unless it is explicitly stated otherwise.
 Due to the commutation relation \eqref{eq:comm} following from the Yang-Baxter relation, the operators
 \(\otimes_i(u_i+2\hD)~\prod_k w(t_k)\) 
are conserved charges, in the sense that they
belong to the same family of commuting operators as the
T-operators. 

The master identity  reads as follows  
(\(t, z \in {\mathbb C}\)): 
\begin{multline}  \label{eq:TQNu} 
(t-z)\lefteqn{\left[\otimes_i(2+u_i+2\hD)~w(z)w(t)\PPi\right] \cdot
  \left[\otimes_i(u_i+2\hD)~\PPi\right]}
\\[4pt]
  = t\left[\otimes_i(u_i+2\hD)~w(\z)\PPi\right] \cdot
  \left[\otimes_i(2+u_i+2\hD)~ 
w(t)\PPi \right]
\\[4pt]
-z \left[\otimes_i(2+u_i+2\hD)~w(\z)\PPi\right]\cdot\left[\otimes_i(u_i+2\hD)~w(t)\PPi\right], 
\end{multline}
and it holds for any function \(\PPi(g)\) of the form \(\PPi(g)=\prod_k
w(t_k)\), or equivalently,
\(\PPi(g)=\det(f(g))=\prod_{j=1}^K f(x_j)\) where \(f(z)\) is an
arbitrary fixed  function 
    analytic in the vicinity of \(z=0\). In
\eqref{eq:TQNu}, the 
dots between consecutive brackets stand for multiplication of
operators acting on the quantum space, and each co-derivative operator
\(\hD\) acts on what lies to its right inside the square brackets.

  The proof of our main identity \eqref{eq:TQNu} is given in
  Appendix \ref{sec:proof-mast-ident}, 
  but it can be easily proved directly, for a few small \(N\)'s, on
  Mathematica.  The identity represents a natural
  generalization\footnote{The identity (4.1) of \cite{Kazakov:2007na}
    corresponds to a particular case when \(\PPi=1\) and \(\forall i, u_i=0\).} of
  the equation (4.1) [equation (20) in the arXiv
      version] in \cite{Kazakov:2007na}.

To conclude this section, let us demonstrate the use of the
master identity  \eqref{eq:TQNu} by deducing from it a particular case
of the Hirota relations for the transfer matrices in particular
representations. For  that purpose, let us focus on the case when
\(\PPi=1\).  
Then, by expending \(w\)'s in symmetric characters according to
\eqref{eq:GenCh} and keeping the coefficient of \(t^s z^{s'}\) in each
term of 
\eqref{eq:TQNu}, we get the relation
\begin{eqnarray}  \label{eq:TQNub} 
\lefteqn{\left[\otimes_i(2+u_i+2\hD)~
    (\chi_{s-1}\chi_{s'}-\chi_{s}\chi_{s'-1})\right]
  \cdot   \left[\prod_i(u_i)\right]}\hspace{50pt}~ \nonumber\\[4pt]
  &=&\left[\otimes_i(u_i+2\hD)~\chi_{s'}\right] \cdot
  \left[\otimes_i(2+u_i+2\hD)~ 
\chi_{s-1}\right]\nonumber\\[4pt]
&&- \left[\otimes_i(2+u_i+2\hD)~\chi_{s'-1}\right]\cdot\left[\otimes_i(u_i+2\hD)~\chi_s\right].
\end{eqnarray}
Choosing \(s'=s+1\), and rewriting \eqref{eq:TQNub}
using\footnote{We also use 
  the fact that the characters for rectangular representations  
\(\chi^{(a,s)} \equiv \chi^{\{(s^a)\}}\), satisfy a simple Hirota relation
  \(\chi^{(a,s+1)}\chi^{(a,s-1)}-\chi^{(a,s)}\chi^{(a,s)}= -
  \chi^{(a-1,s)}\chi^{(a+1,s)}\).} 
\eqref{DIFFT}, we get a relation in terms of the T-operators of
rectangular  Young 
diagrams :
\begin{align}
  -\Tb^{(2,s)}(u+2)\Tb^{(0,s)}(u)=\Tb^{(1,s+1)}(u)\Tb^{(1,s-1)}(u+2)-\Tb^{(1,s)}(u+2)\Tb^{(1,s)}(u)
\label{eq:hir00}
\end{align}
where \((a,s)\) denotes the representation with an \(a\times s\)
rectangular   Young diagram \(\l=(s^a)\), 
i.e. \(\Tb^{(a,s)}(u)=\Tb^{\{(s^a)\}}(u)\).

Eq. \eqref{eq:hir00},  is a particular case of general Hirota
relations ~\eqref{eq:hirotaN} for the fusion in rational spin chains
known since long and proven in \cite{Kazakov:2007na} at zero level of nesting  (\(I=\F\)).

\section{Baxter relations for T- and Q-operators}\label{sec:TQ}

In this section we will derive from our main identity the operatorial
B\"acklund flow in the form of  the  TQ-relations described above, and
even more generally, of TT-relations at every step of the nesting
procedure, as well as the  QQ-relations
\cite{QQ-boson,Bazhanov:2001xm,GS03,Belitsky:2006cp,Kazakov:2007fy,Zabrodin:2007rq,Gromov:2007ky,%
Bazhanov:2008yc}
(see also earlier papers \cite{Woynarovich83}, and 
 a recent presentation in \cite{Tsuboi:2009ud} used in this paper) 
which give 
an immediate access to the full set of  nested Bethe ansatz equations
(also written in an operatorial form in quantum space in the
subsection \ref{subsec:BAE}).  At the same time, it will give a
natural operatorial definition of these quantities on every step of
the nesting, and in particular of all the \(2^{K}\) Q-operators.  
Since all these  T-
and Q-operators belong to the same family of mutually commuting
operators, we can transform these relations, at any stage of the
nesting procedure, to the operatorial ones, for T- and Q-operators.

\subsection{First level of nesting}

\label{sec:Flon}
Now we will obtain from the master  identity \eqref{eq:TQNu} the
operatorial Baxter's TQ-relations. We will start from the first level
of nesting. In what follows we will frequently use the notation \({\overline
  I}=\F \backslash  I\) for the  complimentary set of \(I\). 
 In particular for any element \(j \in \F\), we use a notation 
\({\overline j}=\F \setminus \{j\}\). 

\paragraph{Definition of Q-operators}~

In accord with \eqref{eq:QthroughT} and \eqref{DIFFT} (where \(\Tb^{\{\l\}}(u)\)
denotes \(\Tb^{\{\l\}}_{\overline \emptyset}(u)\)),  the Q-operator on the zero 
level of nesting  is, by definition,
\begin{equation}
  \Qb_{12\dots K}(u)\equiv \Qb_{\overline{\emptyset}}(u)=\left(\prod_{i=1}^{N} u_i\right)\, \mathbb{I}^{\otimes N}
  \label{quantumWron}
\end{equation}
which is a simple function of \(u_i\)'s, times the identity operator in
the full quantum space. 
In particular, the last factor in the l.h.s.
of \eqref{eq:TQNu} becomes \(\Qb_{\overline{\emptyset}}(u)\) when \(\PPi=1\).

We will see in what follows that the Q-operators of the first level of nesting  \(\Qb_{\overline{j}}(u)\)  can be defined through the residues at the poles in the expression:
\begin{equation}
\label{eq:defQN1f}
\left(1-g t\right)^{\bigotimes N}\cdot \left[\bigotimes_i (u_i+2\hat 
    D+2)~w(t) \right] =\sum_{j=1}^{K}\frac{\Qb_{\overline j}(u)}{1-x_j t} +
  \mathrm{polynomial~in~} t , 
\end{equation}
where the normalization factor \(\left(1-g t\right)^{\bigotimes N}\) is
necessary in order to have only simple poles. Indeed, the co-derivative acting on \(w(t)\), having simple poles at each \(t=x_j^{-1},\, j=1,2,\dots,K,\) produces  the double poles in the same points, as it is clearly seen from \eqref{eq:diag2} in the diagonal basis. The factor  \(\left(1-g t\right)^{\bigotimes N} \) transforms them again into simple poles, thus justifying the pole expansion in the r.h.s. of \eqref{eq:defQN1f}.  

The equivalent definition of the Q-operators is 
\begin{equation}
\label{eq:defQN1}
  \Qb_{\overline \jmath}(u)=\lim_{t\to\frac 1
    {x_j}}(1-x_j t) ~ \left(1-g t\right)^{\bigotimes N}\cdot
  \left[\bigotimes_i (2+u_i+  
    2 \hD ) ~ w(t) \right]. 
\end{equation}
This Q-operator acts on the same quantum space
\(\mathcal{H}=(\mathbb{C}^K)^{\otimes N}\) as  the T-operator
\eqref{DIFFT}. 
It is also important to notice that the Q-operator
  \(\Qb_{\overline \jmath}(u)\) looses its dependence on some
  \(u_i\)'s (see appendix \ref{sec:redU}).
For instance, for one spin (\(N=1\)), if we denote\footnote{In this
  definition of the ``diagonal basis'', \(|e_k\rangle\) is such that \(g
  |e_k\rangle= x_k |e_k\rangle\).} by \(|e_k\rangle\)
a basis of eigenstates of \(g\), we get 
\begin{align}
\Qb_{\overline \jmath}(u)|e_k\rangle =\left\{
  \begin{array}{rr}
\left(\left(u_1+2\right)\left(1-x_k/x_j\right)+2{x_k/x_j}\right) \prod_{l\in \overline \jmath}\frac
1 {1-x_l/x_j}|e_k\rangle&\qquad\textrm{if }k\neq j\\[4pt]
2 \prod_{l\in \overline \jmath}\frac
1 {1-x_l/x_j}|e_k\rangle&\textrm{if }k = j , 
\end{array}\right.
\end{align}
 where we see that on the state \(|e_j\rangle\), \(\Qb_{\overline
  \jmath}(u)\) is \(u\)-independent, while the action of other states is
less trivial.

\paragraph{T-operators and TQ-relations}~

Now we will transform the master identity \eqref{eq:TQNu} into a set
of TQ-relations \eqref{eq:TQfun} on the first level of
nesting\footnote{The ``first level of nesting'' means that we will relate
  the original T- and Q-operators labeled by the full set \(\overline
  \emptyset=\F\) with some T and Q-operators labeled by \(\overline j\),
  which has one  index less.}.
 For 
that we simply put \(\PPi=1   \). Multiplying \eqref{eq:TQNu} by the matrix 
\(\left(1-g t\right)^{\bigotimes N}\) which commutes with all the  factors of
both the L.H.S. and the R.H.S.\footnote{Which is clear in the 
diagonal basis
  since \(\left(1-g t\right)^{\bigotimes N}\)
  obviously commutes with permutations, and with tensorial product of
  diagonal matrices, and hence with any operator of the form
  \(\otimes_i(a_i+\hD)~w(b)\), due to its diagrammatic expansion given
  in appendix \ref{sec:diagrammatics}.}, and
picking the poles at \(t=1/x_j\) we come to the equation 
\begin{align}
 \label{eq:TOTQ} 
\begin{split} 
\lim_{t\to\frac{1}{x_j}}&(1-x_j t)(1-z/t) ~ \left(1-g t\right)^{\bigotimes
    N}\cdot\left[\otimes_i(2+u_i+2 \hD)~w(z)w(t)\right] \cdot
\Qb_{\overline\emptyset }(u)=\\ 
  &~~=\left[\otimes_i(u_i+2\hD)~w(\z)\right] \cdot \Qb_{\overline j}(u)
-
x_j \left[\otimes_i(2+u_i+2 \hD)~z~w(\z)\right]\cdot \Qb_{\overline
  j}(u-2) . 
\end{split}
\end{align}
It is useful to note that the factor  \((1-z/t)\sim (1-z x_j)\)
in the L.H.S. 
can be carried over to the right of the 
co-derivatives \(\hD\) allowing to use the relation 
\[(1-z~x_j)w(z)=(1-z~x_j)\det\frac{1}{1-z~g}=\det\frac{1}{1-z~g^{}_{\overline j}}, \] 
where
\( g^{}_{\overline{\jmath}}=\diag(x_1,x_2, \cdots, x_{j-1},x_{j+1},\cdots,x_K) \)
 in the diagonal basis.
 The possibility to move this factor across the derivatives 
comes from the factor \(\left(1-g
  t\right)^{\bigotimes N}\),
introduced to avoid poles of higher orders
in \eqref{eq:defQN1f}. Indeed,  for example in  the
simplest, one spin case  \(N=1\), we can easily check that
\((x_j\I-g)\cdot\hD~ x_j=(x_j\I-g)P_j\cdot x_j=x_j^2P_j-gP_jx_j=0\),
where \(P_j\) is the projector on the
\(j^{\mathrm{th}}\) eigenspace of \(g\). 
The generalization to any \(N\) is rather trivial and is discussed in
the Appendix \ref{sec:Cre}.

 Now  we  introduce the characters of the first level of
 nesting \linebreak \(\chi_s(g^{}_{\overline
   j})=\chi_{\{\lambda=(s)\}}(g^{}_{\overline j})\) corresponding to
 the symmetric tensor representations of the sub-algebra
 \(gl(K-1)\subset gl(K)\), defined by the generating function
\begin{equation}
\label{eq:genbj}
w_{\overline j}(z)\equiv
\det\frac{1}{1-z~g^{}_{\overline{j}}}=
\frac{1}{\prod_{k \in {\overline j}}(1-z~x_{k})}=
\sum_{s=0}^\infty z^s\,\chi_s(g^{}_{\overline j})
\end{equation} 
and define the T-operators of the first level of 
nesting labeled by a Young diagram \(\l=(s)\):
 \begin{equation} 
\label{eq:defTN01}
  \Tb_{\overline j}^{s}(u)=\lim_{t\to\frac 1
    {x_j}} \left(1-x_j ~t\right)\left(1-g~ t\right)^{\bigotimes N}\cdot
  \left[\bigotimes_i   (u_i+2+2 \hat D)  \,\chi_s(g^{}_{\overline j})w(t)\right]. 
\end{equation}
The last formula also  allows for an alternative to 
definition \eqref{eq:defQN1f} of the 
Q-operators as of the T-operators for the singlet irrep in the
auxiliary space corresponding to an empty Young diagram \(s=0\): 
\(  \Qb_{\overline j}(u)\equiv \Tb_{\overline j}^{s=0}(u)\) confirming the (more general) relation announced in \eqref{eq:QthroughT}.

In the definition \eqref{eq:defTN01}, we 
take a residue of a given pole and  at the same time use a character
\(\chi_s(g^{}_{\overline j})\) where one eigenvalue is ``removed''.
This ``removal'' will be at the heart of our nesting procedure, and its repetition
defines a certain  B\"acklund flow (nesting path).

Let us take the coefficient of \(z^{s}\) (\(s \in {\mathbb Z}_{\ge
  0}\)) in \eqref{eq:TOTQ}, analogously to what was done in
\eqref{eq:TQNub}, and rewrite\footnote{As explained above, we also use the
  fact that 
\(w(z)(1-x_j z)=w_{\overline j}(z)\).
} it, using \eqref{eq:defTN01}, as follows:
 \begin{equation}
  \label{eq:TQz}
  \Tb_{\overline{j}}^s(u)\Qb_{\overline{\emptyset }} (u)=
  \Tb_{\overline{\emptyset}}^{s}(u)\Qb_{\overline{j}}(u)-x_j
  \Tb_{\overline{\emptyset}}^{s-1}(u+2)\Qb_{\overline{j}}(u-2) ,
\end{equation} 
which is the simplest Baxter's  TQ-relation in the operatorial form,
the first of the chain of B\"acklund transformations 
among the commuting T- and Q-operators of the zero\(^{\mathrm{th}}\) and first level of
nesting known for a long time
\cite{Krichever:1996qd} on the level of their eigenvalues. 
Here \(\Tb_{\overline{\emptyset}}^{s}(u) \equiv \Tb^{\{(s)\}}(u) \) is the T-operator of the zero\(^{{\mathrm{th}}}\) 
level of nesting, the original transfer matrix \eqref{TMATR}, or
\eqref{DIFFT}, in the symmetric tensor irrep \(\lambda =(s)\). 
The T- and Q-operators
  labeled by \(\overline j=\F\setminus \{j\}\) have \(K-1\)
  indices, and are considered in the first level of nesting\footnote{In
  the same spirit, the \(k^{\mathrm{th}}\) level of nesting will involve the  quantities
  with \(K-k\) indices.}.

Let us also note that the T-operators can be  also defined as the residues at
the poles:
\begin{align}
\label{eq:defTN1p}
  \sum_{j=1}^{K}\frac{t~\Tb_{\overline \jmath}^{s}(u)}{1-x_j t}=& 
\left(1-g t \right)^{\bigotimes N}\cdot \left[\bigotimes_i (u_i+2+2\hat 
    D) \left(t~\chi_{\{s\}}(g)-\chi_{\{s-1\}}(g)\right)  ~w(t)\right]\nonumber\\
&+\mathrm{polynomial~in}~t . 
\end{align} 
It is clear from these definitions and from \eqref{eq:comm}
that all these T- (and hence  the Q-)opera\-tors, belong to the 
same family of commuting operators \([\Tb_{\overline
  \jmath}^{s}(u),\Tb_{\overline\jmath'}^{s'}(u')]=[\Tb_{\overline
  \jmath}^{s}(u),\Tb_{\emptyset}^{s'}(u')]=0.\) It will be also shown
for all T- and Q-operators, on all levels of nesting.

\subsection{Next levels of nesting }

Now we will generalize this procedure, and the corresponding
TQ-relations, to all nesting  levels. Suppose we want to consecutively
``remove''  the eigenvalues \(x_{j_1},x_{j_2},\dots,x_{j_k}\) from the 
characters in the definition of T-operators, where
\(\bI=\{j_1,j_2,\dots,j_{k}\}\) is a subset of the full set of indices:
\(\bI\subset\F\) (their order is not important but they are all
different).    At such  arbitrary  level of nesting, we define a
normalization operator 
\begin{equation}
  \label{eq:QNorm}
  \mathbf{B}_{\bI}
=\prod_{j\in \bI} (1-x_{j}~
t_j)\cdot (1-g~t_j)^{\otimes N}
\end{equation}
and  the following product of generating functions of characters  
\begin{equation}
  {\PPi}_{\bI}=\prod_{j\in \bI} w(t_j) . 
\end{equation}The definition of  the Q-operator labeled by  
 a subset\footnote{the subset \(I\) defines  the node on the Hasse diagram where the nesting process has arrived.}
\(I=\F\setminus \bI\)  
 of the full set \(\F\) 
becomes
\begin{equation}
\label{eq:defQNg}
  \Qb_{I}(u)=\lim_{\substack{t_j\to\frac 1
    {x_j}\\ j\in \bI}}  \mathbf{B}_{\bI}\cdot 
  \left[\bigotimes_i (2 |\bI| +u_i+  
    2 \hD ) ~  \PPi_\bI\right] \qquad \rm{where~~}
   |\bI|=\mathrm{Card}(\bI)=K-|I|
\end{equation}
and once again, it is an operator on the quantum space \((\mathbb{C}^K)^{\otimes N}\), which is polynomial (of degree \(N\) if \(I\neq \emptyset\)) in the
spectral parameter \(u\). Its eigenvalues have degree \(\leq N\), and it
can be shown (see appendix \ref{sec:redU}) that
\(\Qb_I(u)|e_{k_1,k_2,\cdots k_N}\rangle\) is independent of all \(u_n\)
such that \(k_n\in\bI\).    

We will show how to write a \(TQ\) relation between the T- and
Q-operators labeled by \(I\) and the operators of the previous level of
nesting, labeled by \(I\cup j_k\).  

Let us first generalize \eqref{eq:genbj} to define the characters of
\(g^{}_{I}=\diag\left((x_j)_{j\in I}\right)\):
\begin{equation}
\label{eq:chiwI}
w_I(z)\equiv\det\frac{1}{1-z~g^{}_{I}}\equiv 
\prod_{j\in I}\frac 1 {1-z~x_j}\equiv
\sum_{s=0}^\infty z^s\,\chi_s(g^{}_{I}) = \frac{w(z)}{w_{\bI}(z)}.
\end{equation} 
If we chose in the master
identity \eqref{eq:TQNu}, 
 \(\PPi=\PPi_{\overline{I\cup j_k}}\equiv \PPi_{\bI\setminus j_k}\),
 \(t=t_{j_k}\), and \(u\to u+2|\overline{I\cup j_k}|=u+2|\bI|-2\), then after
 multiplying\footnote{
Once again, the normalization factors \(\mathbf{B}_\bI\) and
\(\mathbf{B}_{\overline{I\cup j_k}}\) commute with all the other
factors, because they commute with all permutations, and with all
 operators \(g^{i_1}\otimes g^{i_2}\otimes\cdots g^{i_N}\), which
 are the building blocks of all  other factors.

On the other hand, we will see a posteriori in appendix \ref{sec:Cre}
that the factor \(w_{\overline{I}}(z)\) can be freely move across the
\(\hat D\)'s.
} it by \(\frac 1 {w_{\bI\setminus j_k}(z)}\mathbf{B}_\bI\cdot \mathbf{B}_{\overline{I\cup
     j_k}}\) and taking the limit \(t_j\to \frac 1 {x_j}\), we get
\begin{multline}
\label{eq:MInest}
\hspace{-15pt}
\lim_{\substack{t_j\to\frac 1
    {x_j}\\ j\in \bI\cup j_k}}
\left(
 (1-z /t_{j_k})
 \frac{ \mathbf{B}_{\bI}}
{w_{\overline{I\cup j_k}}(z)}
\cdot \left[\otimes_i(2|\bI|+u_i+2 \hD)~
w(z)w(t_{j_k}) \PPi_{\overline{I\cup j_k}}
\right]
\right) 
\cdot
  \Qb_{I\cup j_k}(u)
\\
=\lim_{\substack{t_j\to\frac 1
    {x_j}\\ j\in \bI}} \left(
  \frac{\mathbf{B}_{\overline{I\cup j_k}}}{{w_{\overline{I\cup
          j_k}}(z)}} \cdot \left[\otimes_i(2|\bI|-2+u_i+2 \hD)~ 
{w(z)}{}\PPi_{\overline{I\cup j_k}}
\right] \cdot
  \Qb_{I}(u)\right.
\\
-x_{j_k} \left. ~
\frac{  \mathbf{B}_{\overline{I\cup j_k}}}
{{w_{\overline{I\cup j_k}}(z)}}
\cdot \left[ \otimes_i(2|\bI|+u_i+2 \hD)~
z {w(z)}\PPi_{\overline{I\cup j_k}}
\right] \cdot
  \Qb_{I}(u-2)\right) ,
\end{multline}
where \(\overline{I\cup j_k}={\{j_1,j_2,\dots,j_{k-1}\}}\).
 These expressions
are obtained by rewriting the z-independent factors using the formula
\eqref{eq:defQNg}. For instance, the last 
factor of the last term obtained from \eqref{eq:TQNu} is 
\begin{align}
\mathbf{B}_{\overline{I}} \cdot\left[\otimes_i(2|\bI|-2+u_i+2
\hD)~w(t_{j_k}) \PPi_{\bI\setminus j_k}\right]
=
\mathbf{B}_{\overline{I}}\cdot\left[ \otimes_i(2|\bI|-2+u_i+2
\hD)~ \PPi_{\bI }\right],
\end{align}
which becomes \(\Qb_{I}(u-2)\) when the limit
\(t_j\to \frac 1 {x_j}\) is taken.

We define the T-operators for symmetric tensor representations as follows
\begin{eqnarray}
  \label{defTs}
\Tb_{I}^s(u)= \lim_{\substack{t_j\to \frac 1 {x_j}\\j\in
    \overline I}}
\mathbf{B}_\bI \cdot \left[\bigotimes_{i=1}^N (u_i+2 \hat D+2 |{\bI}|)\,\,\,\,
    \chi_{s}(g^{}_I) 
 \PPi_\bI\right],
\end{eqnarray}
where \(\chi_{s}(g^{}_I)\) is defined by \eqref{eq:chiwI}. Then the action of co-derivatives  on
\(\chi_{s}(g^{}_I)\) is a priori rather nontrivial. The recipe to avoid this complication and compute T-operators
 is given in appendix \ref{sec:Cre}. In terms of the generating function of \(\Tb_I^s(u)\) we have 
\begin{align}  
{\mathfrak W_I(u,z) \equiv}
 \sum _{s=0}^{\infty}z^s \Tb_{I}^s(u)&= \lim_{\substack{t_j\to \frac 1 {x_j}\\j\in
    \overline I}}
\mathbf{B}_\bI \cdot \left[\bigotimes_{i=1}^N (u_i+2 \hat D+2 |{\bI}|)\,\,\,\,
    \frac{w(z)}{w_{\bI}(z)}
 \PPi_\bI\right] \label{defTc0} \\
&= \frac{1}{w_{\bI}(z)}\lim_{\substack{t_j\to \frac 1 {x_j}\\j\in
    \overline I}}
\mathbf{B}_\bI \cdot \left[\bigotimes_{i=1}^N (u_i+2 \hat D+2 |{\bI}|)\,\,\,\,
   w(z)
 \PPi_\bI\right] . \label{defTc}
\end{align}

Finally, we can notice that the L.H.S. of \eqref{eq:MInest}
contains
 \( \frac{(1-z x_{j_k})}{{w_{\overline{I\cup j_k}}(z)}}
=\frac{1}{w_{\overline I}(z)}\). Then, 
expanding \eqref{eq:MInest} with respect to \(z\) and taking the
coefficients of \(z^{s}\), we obtain using the definition \eqref{defTc}
the following operatorial TQ-relation
 \begin{equation}
  \label{eq:TQg}
  \Tb_{{I}}^{s}(u)\Qb_{I\cup j_k} (u)=
  \Tb_{I\cup j_k}^{s}(u)\Qb_{I}(u)-x_{j_k}
  \Tb_{I\cup j_k}^{s-1}(u+2)\Qb_{I}(u-2) .
\end{equation}
It generalizes the similar, obvious  relation among the characters of symmetric tensor irreps:
\begin{equation}\label{chiBk}
\chi_s(g_{I})=\chi_s(g_{I,j})-x_j
\chi_{s-1}(g_{I,j}),
\end{equation}
where \(j \in \overline{I}\) and \(s \in {\mathbb Z}_{\ge 1}\).

Notice that again
the Q-operator on any level of nesting is equal to the T-operator,
with the same index set \(I,\)  
for an empty Young diagram 
 \(\lambda =\emptyset\) (which corresponds to \(s=0\) case): 
\begin{equation}
\label{eq:deftQfromT}
  \Qb_{I}(u)=\Tb_{I}^{\{\emptyset\}}(u).
\end{equation}

\subsection{Generalization to any representations}

  There is a  natural way to generalize the T-operators to any
  irreps \(\lambda  \) in the auxiliary space: 
\begin{eqnarray}
  \label{defTI}
\Tb_{I}^{\{\lambda\}}(u)= \lim_{\substack{t_j\to \frac 1 {x_j}\\j\in
    \overline I}}\mathbf{B}_{\bI}\cdot
 \left[\bigotimes_{i=1}^N (u_i+2 \hat D+2 |{\bI}|) \, \, 
    \chi_{_{\lambda}}(g^{}_{I}) 
  \PPi_\bI\right], 
\end{eqnarray}
where the  \(gl(K- |{\bI}|)\) characters of the  irreps
\(\lambda\)  are given through the characters of the symmetric tensor representations \(\chi_s(g^{}_I)\)
by the Jacobi-Trudi determinant formula  
\begin{eqnarray}\label{CHADET}
 \chi_{\{\lambda\}}(g^{}_I) = \det_{1\le i,j\le a}\chi_{_{\lambda_j+i-j}} (g^{}_I)\, ,
\end{eqnarray}
where \(a\) is the number of rows in the Young diagram \(\lambda\). 
It is noteworthy that, due to the definition \eqref{defTI},
\(\Tb_I^{\{\l\}}=0\) if \(\l\) has more than \(|I|\) rows, 
because \(\chi_{\{\l\}}(g_I)=0\). 

The Bazhanov-Reshetikhin (BR) formula proven in
\cite{Kazakov:2007na} at the zero\(^{\mathrm{th}}\) level of nesting is also true for    the 
T-operators on every level of nesting, and it reads
\begin{align}
\Tb^{\{ (\l_1,\l_2,\cdots,\l_a) \}}_{I}(u)&=\frac 1
{\prod_{k=1}^{a-1}\Qb_I(u-2k)}
\det_{1\le i,j\le a}\left(\Tb_I^{\l_j+i-j}(u+2-2i)\right) .  
\end{align}
In the particular case of the
rectangular Young diagrams \(\lambda=(s^a)\)  this means
that the T-operators   satisfy the immediate
equivalent of the BR formula -  the Hirota
equation\footnote{Hirota relation can also be written
  in terms of the T-operators defined as  
\(\mathbb T_{a,s}(u)=\Tb^{(a,s)}(u+a-s)\). In term of these \(\mathbb
  T\)-operators, \eqref{eq:hirotaN} takes the usual form
  \(\mathbb T_{a,s}(u+1)\mathbb T_{a,s}(u-1)=\mathbb T_{a+1,s}(u)\mathbb T_{a-1,s}(u)+
\mathbb T_{a,s+1}(u)\mathbb T_{a,s-1}(u)\), more frequent in the
literature.
}, which is the same  on any level of nesting, namely 
\begin{multline}
  \label{eq:hirotaN}
  \Tb_I^{(a,s)}(u+1)\Tb_I^{(a,s)}(u-1)=\Tb_I^{(a+1,s)}(u+1)\Tb_I^{(a-1,s)}(u-1)
\\
+\Tb_I^{(a,s+1)}(u-1)\Tb_I^{(a,s-1)}(u+1).
\end{multline}
 It can be proven that it is  a consequence of the equation
  \eqref{master-det}\footnote{To prove the BR at arbitrary levels of
    nesting, one actually 
  needs to rewrite \eqref{master-det} in the following slightly more
  general form, which is also equivalent to \eqref{eq:TQNu}:
  \begin{align}
    \W_{I,J}(u) &= \frac{\det\left( z_{j}^{n-k}
        \W_{I,j}(u-2k+2) \right)_{\substack{j\in J\\1\leq k\le
          |J|}} } {\det\left(
        z_{j}^{n-k} \right)_{\substack{j\in J\\1\leq k\le
          |J|}} \prod_{k=1}^{|J|-1}
      \W_I(u-2k) } ,
  \end{align}
which becomes the nested Bazhanov-Reshetikhin formula after the limit 
\(z_i\to 1/x_i\) is taken for \(i\in I\).}
    of appendix \ref{sec:proof-mast-ident}.
It is also well known that the TQ-relation \eqref{eq:TQg}
  implies the generating series expression \eqref{eq:series} of
  T-operators for symmetric representations in terms of Q-operators,
  detailed in the appendix 
  \ref{Appx:TQ} (see for instance
  \cite{Krichever:1996qd} at the level of eigenvalues).
Then the Bazhanov-Reshetikhin determinant formula (or, equivalently, the
Hirota equation)  allows to write arbitrary
\(\Tb_{I}^{\{\lambda\}}(u)\) operators in terms of Q-operators and to
check  
the following bilinear relations on the nested T-operators 
\cite{Krichever:1996qd} (see \cite{Zabrodin:2007rq} for the case with non-zero twist \(g\)):
\begin{multline}
\label{eq:BT1}
  \Tb_{I,j}^{(a+1,s)}(u)\Tb_{I}^{(a,s)}(u) -
  \Tb_{I,j}^{(a,s)}(u)\Tb_{I}^{(a+1,s)}(u) 
  \\ 
= x_j \Tb_{I,j}^{(a+1,s-1)}(u+2)
  \Tb_{I}^{(a,s+1)}(u-2), 
\end{multline}
\begin{multline}
\Tb_{I,j}^{(a,s+1)}(u)\Tb_{I}^{(a,s)}(u) -
\Tb_{I,j}^{(a,s)}(u)\Tb_{I}^{(a,s+1)}(u)
\\
=x_j \Tb_{I,j}^{(a+1,s)}(u+2)
  \Tb_{I}^{(a-1,s+1)}(u-2) . 
\label{eq:BT2}
\end{multline}
The TQ-relation \eqref{eq:TQg} is a particular case of equation
\eqref{eq:BT1} when \(a=0\), and the two equations
(\ref{eq:BT1}, \ref{eq:BT2}) coincide with the definition of B\"acklund flow given
in \cite{Zabrodin:2007rq}
up to the permutations on the index set \(I\), so that the definitions
(\ref{eq:deftQfromT}, \ref{defTI})
 for nested
T- and Q-operators explicitly give a solution of the linear
system (\ref{eq:BT1}, \ref{eq:BT2}).

\subsection{QQ-relations}

In the previous subsections, the TQ-relations were proven from the
formula \eqref{eq:TQNu}. This formula contains the explicit factors
\(w(z)\) and \(w(t)\), and one of them was incorporated into \(\PPi\)
(by defining \(\PPi_{\overline{I\cup j_k}}=w(t)\PPi_\bI\)) and was
associated to the nesting level. The other factor (\(w(z)=\sum z^s
\chi_s\)) was decomposed into the characters of \(s\)-symmetric representations.

But it is also  possible to generate other identities from 
\eqref{eq:TQNu}, for instance by incorporating both \(w(t)\) and \(w(z)\) into
\(\PPi_\bI\), which gives rise to different nesting paths.   
Then, a careful rewriting\footnote{In \eqref{eq:TQNu}, let us put 
 \(t=t_{j}, z=t_{i}\), \( \Pi=\Pi_{\overline{I \cup \{i,j\}}}\), 
where \(I \subset \F\) and  \(i,j \in \overline{I}\) (\(i \ne j\)). 
Then note the following relations: \(w(z)w(t) \Pi = \Pi_{\bar{I}}\), 
\(w(z) \Pi = \Pi_{\overline{I \cup \{j\}}}\), 
\(w(t) \Pi = \Pi_{\overline{I \cup \{i\}}} \). 
}
 of \eqref{eq:TQNu} immediately gives the following
well known QQ-relations between Q-operators:
\begin{equation}
\label{eq:QQ}
  \left(x_i-x_j\right) \Qb_{I}(u-2)\Qb_{I,i,j}(u)=  x_i
  \Qb_{I,j}(u-2)\Qb_{I,i}(u) - x_j
  \Qb_{I,j}(u)\Qb_{I,i}(u-2), 
\end{equation}
where \(i,j \in \overline{I}\) and \(i \ne j \).
These are relations among Q-operators in 4-cycles made of 
\(\{\Qb_{I},\Qb_{I,i},\Qb_{I,j},\Qb_{I,i,j}\}\) in the Hasse diagram (cf. figure \ref{Hassegl34}). 

 We can immediately
solve these QQ-relations, which are actually a particular case of the
 Pl\"{u}cker
 identities for all the Q-operators to get the following
determinant representations\footnote{
The factor \(\det\left(x_j^{|J|-1-k}\right)_{\substack{j\in J\\0\leq k\leq |J|-1}}=
\prod_{i <j; i,j \in J}(x_{i}-x_{j})\) in the denominator 
corresponds to the denominator formula of the character of \(gl(|J|)\).
}
\begin{equation}
\label{eq:WronskQ}
  \Qb_{{I,J}}(u)=\frac{\det\left(x_j^{|J|-1-k} \Qb_{{I,j}}(u-2k)
\right)_{\substack{j\in J \\ 0\leq k\leq|J|-1}}}
{\left(\prod_{k=1}^{|J|-1} \Qb_{{I}}(u-2k)  \right)
\det\left(x_j^{|J|-1-k}\right)_{\substack{j\in J\\0\leq k\leq |J|-1}}}.
\end{equation}
In particular, choosing \(I=\emptyset\) gives the
expression of any Q-operator in
terms of \(K+1\) Q-operators, namely, the \(K\) single indexed \(\Qb_{{i}}(u) \)
operators describing the last level of nesting, and the
\(u\)-independent operator \(\Qb_{{\emptyset}}(u)\). More explicitly
\(\Qb_{{\emptyset}}(u)\) can be defined by its action on the diagonal basis of the
quantum space by  
\begin{align}
  \label{eq:TQempty}
 \Qb_{\emptyset}(u)|e\rangle &\equiv
  \Tb^{s=0}_{\emptyset}(u)|e\rangle =
2^N 
\prod_{k=1}^{K} n_{k}! 
\prod_{\substack{j=1, \\ (j\ne k)}}^{K}
\left(
1-\frac{x_{j}}{x_{k}}
\right)^{n_{j}-1}
|e\rangle ,
  \end{align}
where \(|e\rangle =
|e_{i_1,i_2,\cdots,i_N}\rangle\equiv
|e_{i_1}\rangle\otimes|e_{i_2}\rangle\otimes \cdots \otimes|e_{i_{N-1}}\rangle
\otimes|e_{i_N}\rangle \) 
and \(n_{k}\) is the number of \(j \) such that \(i_{j}=k\).


\subsection{Operatorial Bethe equations}\label{subsec:BAE}

In this subsection we derive the set of nested Bethe ansatz equations.

From the QQ-relations  \eqref{eq:QQ} one immediately sees that since
\(\Qb_{I,j}(u)\) should be, by its definition, a polynomial of \(u\),
then (for \(i,j \in \overline{I}\) and \(i \ne j \))
\begin{align}
  \Qb_{I,i}(u) \ &| \ (x_i-x_j)\Qb_{I}(u-2)\Qb_{I,i,j}(u)+x_j
  \Qb_{I,j}(u)\Qb_{I,i}(u-2), \\[6pt]
\Qb_{I,i}(u) \ &| \ (x_i-x_j)\Qb_{I}(u)\Qb_{I,i,j}(u+2)-x_i
  \Qb_{I,j}(u)\Qb_{I,i}(u+2)  ,
\end{align}
where \(P|P'\) denotes the fact that the polynomial \(P'\) contains the polynomial \(P\) as a factor. 
By adding \(x_i \Qb_{I,i}(u+2)\) times the first line to \(x_j
\Qb_{I,i}(u-2)\) times the second line, one gets
\begin{align}\label{eq:OpBAE}
  \Qb_{I,i}(u) \ &| \ 
 x_i\Qb_{I}(u-2)\Qb_{I,i,j}(u)\Qb_{I,i}(u+2) 
+ 
x_j \Qb_{I}(u)\Qb_{I,i,j}(u+2)\Qb_{I,i}(u-2).
\end{align}
This is written for the Q-operators, but 
when acting on particular eigenstates, the operatorial  Bethe
equations \eqref{eq:OpBAE}  
become the usual polynomial Bethe equations \eqref{BAE-eigen} 
\cite{BAE-ref,Kulish:1983rd} 
(cf.\ eq.\ (68) of \cite{Zabrodin:2007rq})
 on the Bethe roots
\( \{u^{(I)}_k \} \) along a chosen nesting path. 
 Indeed, now we know,
by construction, that the Q-operators
are polynomials\footnote{This fact was missing in the analytic Bethe
  ansatz construction of 
  \cite{Zabrodin:2007rq}  and appeared there only as a hypothetic
  ansatz for the solution of Hirota equation by the B\"acklund
  procedure.} of \(u\) 
and therefore for each eigenstate their eigenvalues are also
polynomials of a degree \(K_I\leq N\) in \(u\): 
\begin{equation}
{{\mathsf Q}_{I}(u)=c_I\prod_{k=1}^{K_I}(u-u^{(I)}_k)},
 \label{Q-function}
\end{equation} 
 so that, substituting \(u=u^{(I,i)}_{k}\) into an eigenvalue of \eqref{eq:OpBAE}, we obtain the usual nested Bethe ansatz equations:
\begin{align}
 -1=\frac{x_{i}}{x_{j}} 
 \frac{\Qs_{I}(u^{(I,i)}_{k}-2)\Qs_{I,i}(u^{(I,i)}_{k}+2)\Qs_{I,i,j}(u^{(I,i)}_{k})}
{\Qs_{I}(u^{(I,i)}_{k})\Qs_{I,i}(u^{(I,i)}_{k}-2)\Qs_{I,i,j}(u^{(I,i)}_{k}+2)}
\qquad \text{for} \quad 
k \in \{1,2,\dots, K_{I,i}\}.
 \label{BAE-eigen}
\end{align}
The Q-operators are formal polynomials \(\Qb_{I}(u)=\sum_{k=1}^{N}u^k\,\,{\mathbf 
  c}^{(I)}_k\) whose coefficients \(\{{\mathbf c}^{(I)}_k\}\) are
also operators in the quantum space. On particular quantum states - the
eigenvectors of the spin chain Hamiltonian, a part of them becomes zero, which
explains the fact that the power of Q-functions - the eigenvalues of
the Q-operators - can diminish on each step of the B\"acklund
procedure. 

 Given all solutions of the  Bethe equations \eqref{BAE-eigen} it is possible, in principle, to
find all eigenvalues of the \(\Qb_{I_j}\) operators and then
to reconstruct all T-operators  using TQ-relations (see
  \eqref{eq:series}) together with the Hirota relation.

\section{Generalization to the supersymmetric case}\label{sec:super} 

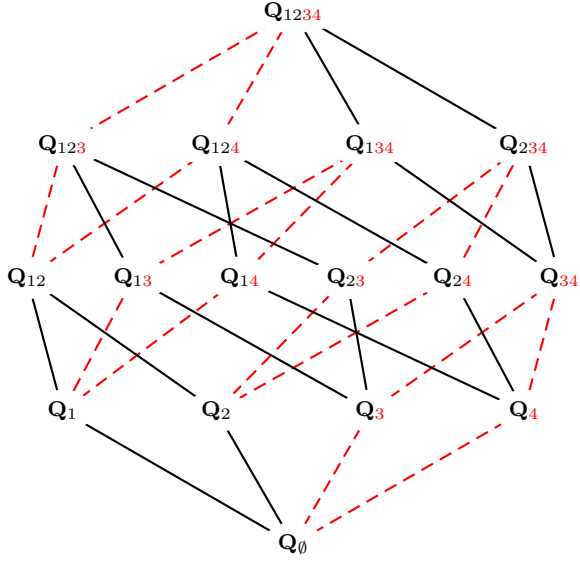
\begin{figure}
  \centering
\psset{unit=1pt}
\begin{pspicture}(0,0)(200,200)
\psline(100,0)(13.4,50)
\psline(100,0)(71.1,50)
\psline[linecolor=red,linestyle=dashed,linewidth=0.8](100,0)(128.9,50)
\psline[linecolor=red,linestyle=dashed,linewidth=0.8](100,0)(186.6,50)
\psline(13.4,50)(0,100)
\psline[linecolor=red,linestyle=dashed,linewidth=0.8](13.4,50)(40,100)
\psline[linecolor=red,linestyle=dashed,linewidth=0.8](13.4,50)(80,100)
\psline(71.1,50)(0,100)
\psline[linecolor=red,linestyle=dashed,linewidth=0.8](71.1,50)(120,100)
\psline[linecolor=red,linestyle=dashed,linewidth=0.8](71.1,50)(160,100)
\psline(128.9,50)(40,100)
\psline(128.9,50)(120,100)
\psline[linecolor=red,linestyle=dashed,linewidth=0.8](128.9,50)(200,100)
\psline(186.6,50)(80,100)
\psline(186.6,50)(160,100)
\psline[linecolor=red,linestyle=dashed,linewidth=0.8](186.6,50)(200,100)
\psline[linecolor=red,linestyle=dashed,linewidth=0.8](0,100)(13.4,150)
\psline[linecolor=red,linestyle=dashed,linewidth=0.8](0,100)(71.1,150)
\psline(40,100)(13.4,150)
\psline[linecolor=red,linestyle=dashed,linewidth=0.8](40,100)(128.9,150)
\psline(80,100)(71.1,150)
\psline[linecolor=red,linestyle=dashed,linewidth=0.8](80,100)(128.9,150)
\psline(120,100)(13.4,150)
\psline[linecolor=red,linestyle=dashed,linewidth=0.8](120,100)(186.6,150)
\psline(160,100)(71.1,150)
\psline[linecolor=red,linestyle=dashed,linewidth=0.8](160,100)(186.6,150)
\psline(200,100)(128.9,150)
\psline(200,100)(186.6,150)
\psline[linecolor=red,linestyle=dashed,linewidth=0.8](13.4,150)(100,200)
\psline[linecolor=red,linestyle=dashed,linewidth=0.8](71.1,150)(100,200)
\psline(128.9,150)(100,200)
\psline(186.6,150)(100,200)
\rput[c]{0}(100,0){\psovalbox*{\(\Qb_{\emptyset}\)}}
\rput[c]{0}(13.4,50){\psovalbox*{\(\Qb_{1}\)}}
\rput[c]{0}(71.1,50){\psovalbox*{\(\Qb_{2}\)}}
\rput[c]{0}(128.9,50){\psovalbox*{\(\Qb_{\color{red}3}\)}}
\rput[c]{0}(186.6,50){\psovalbox*{\(\Qb_{\color{red}4}\)}}
\rput[c]{0}(0,100){\psovalbox*{\(\Qb_{12}\)}}
\rput[c]{0}(40,100){\psovalbox*{\(\Qb_{1{\color{red}3}}\)}}
\rput[c]{0}(80,100){\psovalbox*{\(\Qb_{1{\color{red}4}}\)}}
\rput[c]{0}(120,100){\psovalbox*{\(\Qb_{2 {\color{red}3}}\)}}
\rput[c]{0}(160,100){\psovalbox*{\(\Qb_{2 {\color{red}4}}\)}}
\rput[c]{0}(200,100){\psovalbox*{\(\Qb_{\color{red} 34}\)}}
\rput[c]{0}(13.4,150){\psovalbox*{\(\Qb_{12{\color{red}3}}\)}}
\rput[c]{0}(71.1,150){\psovalbox*{\(\Qb_{12{\color{red}4}}\)}}
\rput[c]{0}(128.9,150){\psovalbox*{\(\Qb_{1 {\color{red}34}}\)}}
\rput[c]{0}(186.6,150){\psovalbox*{\(\Qb_{2 {\color{red}34}}\)}}
\rput[c]{0}(100,200){\psovalbox*{\(\Qb_{12 {\color{red}34}}\)}}
\end{pspicture}

 \caption{Hasse diagram for the Q-operators: \(gl(2|2)\) case. 
 There are \(2^4=16\) Q-operators in the same way as \(gl(4)\) case 
(figure \ref{Hassegl4}). 
\(\Qb_{I}\) and \(\Qb_{I \setminus \{k\}}\) are connected by a 
solid line if \(k \in \{1,2\}\) (bosonic \(p_{k}=0\)), 
dashed line if  \(k \in \{3,4\}\) (fermionic \(p_{k}=1\)). 
}
  \label{Hassegl22}
\end{figure}

In the case of the \(gl(K|M)\) super-spin chain, T- and Q-operators are labelled by the \(2^{K+M}\)
 subsets \(I\) of the full set \(\F=\{1,2,\dots, K+M\}\). 
For any element of \(\F\), we define the grading parameter:
\begin{align}
  \label{eq:defGrad}
  p_b=0 \quad \text{for} \quad 1\leq b\leq K, 
\qquad \text{and} \qquad  p_f=1 \quad \text{for} \quad K+1\leq f\leq K+M .
\end{align}
Now the Q-operators are described by the 
colored Hasse diagram (see figure \ref{Hassegl22}). 

In the case of the \(gl(K|M)\) super-spin chain, the co-derivative can
be defined by 
\begin{equation}
\hat D \otimes f(g)=\sum_{ij} e_{ij}  \frac{\partial}{\partial
  \phi_i^{\,j}} \otimes 
f\left(e^{ \sum_{kl} \phi^k_{\,l} e_{kl}} g\right)_{\phi=0}\,\,\, , \,\,\, \qquad 
\frac{\partial}{\partial \phi_{i_1}^{\,j_1}}\phi_{\,j_2}^{i_2} \equiv
\delta_{j_1}^{i_2} \delta_{j_2}^{i_1} (-1)^{p_{j_1}}, 
\end{equation}
where \(\{e_{ij} \}\) are 
 generators of \(gl(K|M)\) in the fundamental representation and 
 a matrix \(\phi\) is expressed as \(\phi =\sum \phi^{i}_{j}e_{ij}\), and 
 \(g\) is a matrix in the fundamental 
representation of \( GL(K|M) \). 

As explained in \cite{Kazakov:2007na}, the properties of
co-derivatives are exactly the same as in the bosonic case, including
the expression of T-operators in terms of co-derivatives (at the
zero\(^{\mathrm{th}}\) nesting level). The diagrammatics  of the
co-derivative is also the same as in the bosonic case, except the
signs   to be introduced into every permutation operator, to get
\(\P = \sum_{\a\b}(-1)^{p_\a}e_{\b\a}  \otimes  e_{\a\b}\).
In particular, the formula \eqref{eq:TQNu} still holds in the
super-case but \eqref{eq:GenCh} has to be substituted by
\begin{equation}
\label{eq:GensCh}
w(z)=\sdet\frac{1}{1-z~g}=\frac{\prod_{j=1}^M(1-z ~y_j)}{\prod_{j=1}^K(1-z ~x_j)}
=\sum_{s=0}^\infty z^s~\chi_s(g)=\frac{1}{\sum_{a=0}^\infty (-1)^a z^a~\chi^{(a)}(g)},
\end{equation}
where 
\( (x_1,\cdots, x_K, y_{1},\cdots,y_{M}) \equiv 
\left(\xi_1,\cdots,\xi_{K+M}\right) \) 
are the eigenvalues of \(g \in GL(K|M)\) in 
the fundamental representation and
\(\sdet\) denotes the
 super-determi\-nant. 
With slight generalizations of the definitions w.r.t. the bosonic case, all the supersymmetric TQ- and QQ-relations follow from
\eqref{eq:TQNu} if we define T- and Q-operators in the following
way: 
\begin{align}
  {\PPi}_{\bI}&=\prod_{j\in \bI} w(t_j)^{((-1)^{p_j})},
&  \mathbf{B}_{\bI}&=\prod_{j\in \bI} (1-\xi_{j}~ t_j)
(1-g~t_j)^{\otimes N} , 
\label{eq:DefSupB}
\end{align}
\begin{align}
\Tb_{I}^{\{\lambda\}}(u)&=
{\lim_{\substack{t_j\to \frac 1 {\xi_j} \\j\in
    \overline I}}\mathbf{B}_{\bI}\cdot
 \left[\bigotimes_{i=1}^N (u_i+2 \hat D+2 n_{\bar{b}}-2n_{\bar{f}})\,\,\,\,
    \chi_{_{\lambda}}(g^{}_{I}) 
  \PPi_\bI\right],}  
\label{eq:DefSupT}
\\
\Qb_{I}(u)&= 
{\lim_{\substack{t_j\to \frac 1 {\xi_j}\\j\in
    \overline I}}\mathbf{B}_{\bI}\cdot
 \left[\bigotimes_{i=1}^N (u_i+2 \hat D+2 n_{\bar{b}}-2n_{\bar{f}})\,\,\,\,
    \PPi_\bI\right],}
\label{eq:DefSupQ}
\end{align}
where 
\(n_{\bar{b}}=|\bar{I}\cap \{1,2,\cdots,K\}|\), and\footnote{As before, \(|{
  I}|\) denotes \(\mathrm{Card}(I)\)} 
\(n_{\bar{f}}=|\bar{I}\cap \{K+1,\cdots,K+M\}|\). 
In particular, the Q-operator for an empty set \eqref{eq:TQempty} now becomes 
\begin{align}
  \label{eq:TQempty2}
 \Qb_{\emptyset}(u)|e\rangle &\equiv
  \Tb^{\{\emptyset \}}_{\emptyset}(u)|e\rangle =
2^N 
\prod_{k=1}^{K+M} (-1)^{p_{k}n_{k}}n_{k}! 
\prod_{\substack{j=1, \\ (j\ne k)}}^{K+M}
\left(
1-\frac{\xi_{j}}{\xi_{k}}
\right)^{n_{j}-(-1)^{p_{j}+p_{k}}}
|e\rangle .
  \end{align}
In what follows, the indices \(i,j\in\{1,2,\dots,K\} \cap \overline{I}\)
and \(\hat k, \hat l\in \{K+1,K+2,\dots, K+M\} \cap \bar{I}\) will correspond to the opposite gradings (we might call \(i\) and \(j\) ``bosons'' and
\(k,l\) ``fermions''), and we will use the notation
\(\hat{l}=l+K\) for \(l \in \{1,2,\dots, M \}\), so that \(\xi_{\hat{l}}=y_{l}\). 
In \eqref{eq:DefSupT}, \(\chi_\l(g_I)\) is defined through
\eqref{CHADET}, where \(\chi_s(g_I)\) is defined by the generating
series
 \begin{equation}
\label{eq:chiwIs}
w_I(z)\equiv\sdet\frac{1}{1-z~g^{}_{I}}
=\frac{\prod_{\substack{\hat l\in I\\\hat l > K}}(1-z~
  y_l)}{\prod_{\substack{j\in I\\j\leq K}}(1-z~ x_j)}
=\sum_{s=0}^\infty z^s~\chi_s(g_I).
\end{equation} 
That implies  \cite{DM92,Tsuboi:1997iq} that \(\Tb_\F^{\{\l\}}=0\)
if
the Young diagram \(\l=(\l_1,\l_2,\l_3\ldots)\) contains the highest weight \(\l_{K+1}>M\)  
(i.e. unless \(\l\) is inside the ``fat hook region'' indicated in figure 10 in \cite{Kazakov:2007na}),
while for nested T-operators \(\Tb_I^{\{\l\}}\), the same ``fat-hook 
  condition'' holds, but the corner of the fat hook is displaced from
  \((M,K)\) to \((n_f,n_b)\), where  
\(n_b=|I\cap \{1,2,\cdots,K\}|,\quad n_f=|I\cap \{K+1,\cdots,K+M\}|\). 
This means that the ``hook region''
  decreases by one row or one column at each level of nesting.

Then the TQ relations \eqref{eq:TQg}\footnote{
obtained from the master identity \eqref{eq:TQNu} by putting 
\(\Pi=\Pi_{\overline{I \cup \{j\}}}, t=t_{j} \) for \eqref{eq:TQb}; 
\(\Pi=\Pi_{\bar{I}}, t=t_{\hat{j}} \) for \eqref{eq:TQf}. 
} become
\begin{align}
  \label{eq:TQb}
  \Tb_{I}^{s}(u)\Qb_{I,j} (u)&=
  \Tb_{I,j}^{s}(u)\Qb_{I}(u)-x_j
  \Tb_{I,j}^{s-1}(u+2)\Qb_{I}(u-2) 
\\
  \Tb_{I,\hat{l}}^{s}(u)\Qb_{I}(u)&=\Tb^{s}_{I}(u)\Qb_{I,\hat{l}}(u) - y_{l}
  \Tb_{I}^{s-1}(u+2) \Qb_{I,\hat{l}} (u-2) 
\label{eq:TQf}
  \\
\nonumber \lefteqn{\textrm{for } 1 \le j\le K
  \textrm{ and }K+1 \le \hat l  \le K+M}~~~
\end{align}
The QQ-relations\footnote{
obtained from the master identity \eqref{eq:TQNu} by putting 
\(\Pi=\Pi_{\overline{I \cup \{i,j\}}}, t=t_{j},z=t_{i} \) for \eqref{eq:QQbb};
\(\Pi=\Pi_{\overline{I \cup \{i\}}}, t=t_{\hat{l}},z=t_{i} \) for \eqref{eq:QQbf};
\(\Pi=\Pi_{\overline{I}}, t=t_{\hat{m}},z=t_{\hat{l}} \) for \eqref{eq:QQb3}.
} also become grading-dependent:
\begin{align}
\label{eq:QQbb}
 \left(x_i-x_j\right) \Qb_{I}(u-2) \Qb_{I,i,j}(u)&= x_i
  \Qb_{I,j}(u-2)\Qb_{I,i}(u) - x_j
  \Qb_{I,j}(u)\Qb_{I,i}(u-2),\\[6pt]
\label{eq:QQbf}
 \left(x_i-y_l\right) \Qb_{I,\hat{l}}(u-2)\Qb_{I,i}(u) &= \x_i
  \Qb_{{I}}(u-2)\Qb_{{I,i,\hat{l}}}(u)-y_l
  \Qb_{{I}}(u)\Qb_{{I,i,\hat{l}}}(u-2),\\[6pt]
\label{eq:QQb3}
 \left(y_{l}-y_{m}\right) \Qb_{I,\hat{l},\hat{m}}(u-2) \Qb_{I}(u) &= y_{l}
  \Qb_{I,\hat{l}}(u-2)\Qb_{I,\hat{m}}(u) - y_{m}
  \Qb_{I,\hat{l}}(u)\Qb_{I,\hat{m}}(u-2),\\[6pt]
\mathrm{for} \quad  i,j \in \{1,2,\dots, K\}   \cap \bar{I} & 
\quad \mathrm{and} \quad \hat{l},\hat{m} \in \{K+1,K+2,\dots, K+M\} \cap \bar{I}. \nonumber 
\end{align}
These are relations among Q-operators in 4-cycles made of 
\(\{\Qb_{I},\Qb_{I,i},\Qb_{I,j},\Qb_{I,i,j}\}\) in the Hasse diagram fig.~\ref{Hassegl22}. 
For example, in the figure \ref{Hassegl22}, 
eq.\eqref{eq:QQbb} corresponds to   4-cycles made of 4-solid lines, 
\eqref{eq:QQbf} corresponds to 4-cycles made of 2-solid lines and 2-dashed lines, 
\eqref{eq:QQb3} corresponds to 4-cycles made of 4-dashed lines. 

All these TQ- and QQ- relations are derived by choosing an appropriate
\(\PPi\) in \eqref{eq:TQNu}, but, as explained in \cite{Gromov:2010km},
they could have been obtained from the bosonic relations by the so called
``bosonization trick'' : For instance, \eqref{eq:QQbf} can be
rewritten as  
\begin{align} 
\left(x_i-y_l\right)  {\Qb_{J}(u-2)}\linebreak[1]  {\Qb_{J,i\setminus \hat
l}(u)} = \x_i
  \Qb_{{J\setminus \hat l}}(u-2)\Qb_{{J,i}}(u)-y_l
  \Qb_{{J \setminus \hat l}}(u)\Qb_{{J}}(u-2),
\end{align}
 where
  \(J=I\cup\{\hat l\}\), which has exactly the same form as
  \eqref{eq:QQbb}, up the the formal replacement \(J\to I\), \(J\setminus
  \hat l\to I,l\). This is interpreted as the fact that adding a
  ``boson'' 
  to the set \(I\) (which indexes Q or T-operators) is equivalent to
  removing a ``fermion''\footnote{As explained in
    \cite{Zabrodin:2007rq}, the same linear system
describes the addition of a column to the ``fat hook'' or the removal
of a line (see (41) and (42) in \cite{Kazakov:2007fy,Zabrodin:2007rq}). In our
notation, the former case is the addition of a 
``fermion'' to the set I, while the latter is the removal of a ``boson''.
}. This trick can be viewed as a mnemonic
rule, and in this construction of the T- and Q-operators, it
comes\footnote{In 
  (\ref{eq:DefSupB}-\ref{eq:DefSupQ}) we can see that adding a
  ``boson'' to the set \(I\) multiplies \(\Pi_{\bI}\) by \(w(t_j)\), while
  adding a ``fermion'' divides it by \(w(t_j)\). As the ``Master
  Identity'' is only sensible to the addition/removal of \(w(t_j)\)
  factors, it is not surprising that the QQ-relation obtained from
  ``master identity'' satisfies this ``bosonization trick.''} 
  from the power
\((-1)^{p_j}\) in the definition \eqref{eq:DefSupB} of \({\PPi}_{\bI}\).
This trick means
that the Hasse diagram for \(gl(K|M)\) can be ``rotated'' by putting
\(\Qb_{K+1,K+2,\cdots ,K+M}\) on the top row and 
\(\Qb_{1,2,\cdots ,K} \) 
on the bottom row, so that the QQ-relations take the same
form for all the facets of the modified diagram.

In the same way as for the \(gl(K)\) case, the TQ-relations
  (\ref{eq:TQb}, \ref{eq:TQf}) can be written as the generating series
  expansion \eqref{eq:seriesS} (see
  \cite{Kazakov:2007fy,Zabrodin:2007rq} at the  level of eigenvalues). 
We can also generalize the Wronskian expressions \eqref{eq:WronskQ} for
Q-operators to all the T-operators at all levels of nesting. They do
not differ in the form from the relations found in
\cite{Tsuboi:2009ud}.

The Bethe ansatz equations
for \(gl(K|M)\) \cite{BAE-ref}
are obtained in the same way as for the \(gl(K)\) case:
there are two ``bosonic'' Bethe Ansatz equations (BAEs) and two ``fermionic''
ones\footnote{We will call ``bosonic'' (resp ``fermionic'') the Bethe
  Ansatz Equations 
  having two free indices of same grading (resp of opposite grading).
For instance, in \eqref{opBAEff}, the two free indices \(\hat l\) and
\(\hat m\) have the same grading, hence this Bethe equation is called
``bosonic''.}. The first ``bosonic'' BAE follows from \eqref{eq:QQbb} and is
unchanged with respect to the section \ref{subsec:BAE} : 
eq.\ \eqref{eq:OpBAE} on the level of the operator and 
eq.\ \eqref{BAE-eigen} on the level of the eigenvalue. 
The second bosonic BAE is 
obtained by isolating \(\Qb_{I,\hat l}\)  in \eqref{eq:QQb3} 
which gives
\begin{align}
  \Qb_{I,\hat{l}}(u) \ | \ 
y_m\Qb_{I,\hat{l},\hat{m}}(u-2)\Qb_{I}(u)\Qb_{I,\hat{l}}(u+2)
+y_l \Qb_{I,\hat{l},\hat{m}}(u)\Qb_{I}(u+2)\Qb_{I,\hat{l}}(u-2),
 \label{opBAEff}
\end{align}
which is, at the level of eigenvalues, the equation \eqref{BAEff},
equivalent to the equation (69) of
\cite{Zabrodin:2007rq} (up to the permutation on the indices
(the Weyl group symmetry)). 

On the other hand, the ``fermionic'' BAEs can be immediately
obtained from \eqref{eq:QQbf} in the form
\begin{eqnarray}
\label{eq:OBbf}
  \Qb_{I,i}(u)&|&\x_i
  \Qb_{{I}}(u-2)\Qb_{{I,i,\hat{l}}}(u)-y_l
  \Qb_{{I}}(u)\Qb_{{I,i,\hat{l}}}(u-2),\\[6pt]
\label{eq:OBfb}
  \Qb_{I,\hat{l}}(u)&|&\x_i
  \Qb_{{I}}(u)\Qb_{{I,i,\hat{l}}}(u+2)-y_l
  \Qb_{{I}}(u+2)\Qb_{{I,i,\hat{l}}}(u).
\end{eqnarray}
In terms of eigenvalues, the Bethe equation \eqref{eq:OBbf} (resp
\eqref{eq:OBfb}) is written as \eqref{BAEbf} (resp \eqref{BAEfb})
 in the appendix \ref{sec:super-case-bethe}.

\section{Conclusions}

 The co-derivative formalism and the master identity
 \eqref{eq:TQNu}, together with the definitions
   \eqref{eq:defQNg},\eqref{defTI}  of
 nested T- and Q-operators proposed  in this paper can serve as  an
 alternative approach to the quantum integrability, rather different
 from the popular algebraic Bethe ansatz (see for example \cite{Faddeev:1996iy} and the references therein). It allows to complete the
 whole procedure of  diagonalization of  transfer-matrix of the
 inhomogeneous twisted \(gl(K|M)\) (super)spin chain, all the way from 
 its construction from R-matrices obeying the Yang-Baxter relations
 and till the nested system of Bethe ansatz equations, directly in
 terms of the  operators acting on the quantum space. The master
 identity \eqref{eq:TQNu} presented at the beginning of the paper and
 generalizing a similar identity from \cite{Kazakov:2007na} is the
 basis of this approach, encoding all possible operatorial QQ- and
 Baxter's TQ-relations at every step of nesting, or of the operatorial
 B\"acklund flow, generalizing the operatorial B\"acklund
 transformations of
 \cite{Krichever:1996qd,Kazakov:2007fy}. Remarkably, the master
 identity takes a bilinear  form with respect to the \(gl(K|M)\)
 characters, or their generating functions. Since the characters can
 be viewed as the tau-functions of KdV hierarchy (which is of course a reduction of KP hierarchy) one can speculate
 that this identity is simply a particular case of the general Hirota
 identity for the KP tau-functions, with \(\tau_n=\frac{1}{n}\tr
 g^n\) playing the role of the KP ``times''. It would be an
 interesting relation between the quantum and classical integrability,
 showing that, paradoxically, the former is a particular case of the
 latter.

It would be also interesting to generalize our approach to the case of
non-compact representations of \(gl(K|M)\) in the auxiliary space,
following the observations made in
\cite{Gromov:2010vb,Gromov:2010km,Hegedus:2009ky} for the characters
and Q-operators for \(U(2,2|4)\). This might teach us how to deal with
one of the most interesting integrable physical systems, \({\cal N}=4\)
SYM theory and its AdS dual - the Metsaev-Tseytlin sigma-model having
the \(PSU(2,2|4)\) global symmetry. In general, the Y-systems for
sigma-models and their Wronskian solutions
\cite{Gromov:2008gj,Kazakov:2010kf} might be also an interesting
subject for their operatorial generalization in the quantum (physical)
space and might give us an interesting tool for the study of the
spectrum of excited states and shed some light on the formulas for
the energy of an excited state conjectured in the literature for
relativistic sigma models \cite{Bazhanov:1996aq,Dorey:1996re,Kazakov:2010kf} and for the
AdS/CFT \cite{Gromov:2009zb}. It would be also interesting to generalize our master identity to other    symmetries of the spin chains, where the Bazhanov-Reshetikhin-type relations are also known, to other than fundamental irreps  in the quantum space, as well as to the trigonometric and elliptic R-matrices.

\section*{Acknowledgments}
The work  of VK was partly supported by  the grant RFFI 08-02-00287. The work  of VK was also partly supported by  the ANR grant   GranMA (BLAN-08-1-313695). We  thank
 A. Alexandrov, V. Bazhanov, L. F\'evrier, N. Gromov, T. Lukowski,
 C. Meneghelli, 
 M. Staudacher, P. Vieira and especially A. Zabrodin for useful comments and
 discussions.
The work of ZT was
supported by Nishina Memorial Foundation and by
Grant-in-Aid for Young Scientists, B \#19740244 from
The Ministry of Education, Culture, Sports, Science and Technology in Japan.
ZT thanks \'Ecole Normale Superieure, LPT, where a considerable part of this work was done,
for the kind hospitality.

\section*{Note added}

When  finishing this paper we learned about the  results of \cite{BFLMS10} which deal with the same problem, the operatorial formulation of the Q-operators and of the TQ-relations. The objects studied in that forthcoming paper are the same, but the formalism is radically different from ours.

\section*{Notes for version 2}

We  corrected in this version 
[arXiv:1010.4022v2 [math-ph]] 
many minor misprints of the first version, and 
added some details and explanations. 
It is precised that  
\(\Pi(g)\) in the master identity \eqref{eq:TQNu} 
has a form of \(\det f(g)\), where \(f(z)\) is an arbitrary fixed function, rather than an arbitrary class function of the twist matrix \(g\). We  also removed a false TT-relation (eq.~(4.27) in the version 1). 

 Most importantly, we give in this new version a concise proof of our master identity - the basis of our approach.

\section*{Notes for version 3}

Minor misprints are corrected.

\appendix

\section{Diagrammatics of co-derivatives}
\label{sec:Dia}

\subsection{Co-Derivatives and characters}

\label{sec:diagrammatics}
The action of the co-derivative on
characters and their generating function is explained in \cite{Kazakov:2007na}. 
For instance, we can write 
\begin{equation}
\hat D\otimes \hat D ~~ w(\x)=\frac {g \x}{1- g \x} \otimes \frac {g \x}{1- g  \x}+\P_{1,2}\left(\frac {1}{1- g \x} \otimes \frac {g \x}{1- g  \x}\right)= 
\left(\eG{{\VL{blue}\VL{blue}}} + \eG{\DC{blue}}\right)w(\x),\label{eq:diag2}
\end{equation}
where \(\eG{\VL{blue}}\) (resp \(\eG{\dVL{blue}}\)) stands for \(\frac {g \x}{1- g
  \x}\) (resp \(\frac {1}{1- g\x}\)), so that
\(\eG{\VL{blue}\VL{blue}}=\frac {g \x}{1- g \x} \otimes \frac {g \x}{1- g
  \x}\) and \(\eG{\DC{blue}}=\P_{1,2}(\frac {1}{1- g \x} \otimes \frac
{g \x}{1- g \x})\), where \(\P_{1,2}\) denotes the usual
permutation between the \(1^{\mathrm{st}}\) and \(2^{\mathrm{nd}}\)
quantum space\footnote{For the supergroups,
  \(\P_{1,2}\) is replaced by   the super
  permutation which differs from the usual permutation only by certain  signs (see \cite{Kazakov:2007na}).}. The general expression for \(\hat
D^{\otimes N} w(\x)\) is given 
by the formula (4.11) of \cite{Kazakov:2007na} (formula (30)
  in the arXiv version), and it represents the sum of 
diagrams corresponding to all possible permutations, with dashing all
the lines going from  lower  to upper nodes and  directed   to the right.

In terms of characters of the symmetric tensor irreps \(\chi_s(g)\), 
the  equation \eqref{eq:diag2} reads\\ \(\hat D\otimes
\hat D ~ \chi_\t = \left(\eG{\VL{blue}\VL{blue}} +
  \eG{\DC{blue}}\right)\chi_\t\) where \(\eG {\VL{blue}}\) (resp \(\eG{\dVL{blue}}\)) stands for \(\frac {g e^{-\partial_\t}}{1- g
  e^{-\partial_\t}}\) (resp \(\frac {1}{1- g e^{-\partial_\t}}\)). If
we identify \(\x=e^{-\partial_\t}\) there is no ambiguity between these
two definitions of the diagrammatics.

This diagrammatics can also be extended for the inclusion of the parameters \(u_i\), i.e.
\begin{equation}
(u_1 +\hat D)\otimes (u_2 +\hat D) ~w(\x)=
\left(\DVL\DVL+\VL{blue}\DVL+\DVL\VL{blue}+\VL{blue}\VL{blue}+ \DC{blue}\right)w(\x),\label{eq:diag3}
\end{equation}
where \(\DVL\) stands for \(u_1\) or \(u_2\), according to its position (for
instance, \(\DVL\VL{blue}=u_1 \I\otimes \frac {g \x}{1-g \x}\), and
\(\VL{blue}\DVL=\frac {g \x}{1-g \x}\otimes \I u_2\)).

\subsection{Co-derivatives of products}

\label{sec:diagProd}
In this paper, we often use the co-derivatives acting on products,
defining the  quantities like \(\hat D^{\otimes N}~\PPi \cdot \CF \)
for an arbitrary class 
  function \(\PPi\). Then each co-derivative can act either on \(\PPi\) or
  on \(\CF\), and at two spins, for instance, the Leibniz rule gives 
  \begin{align}
    \left[\hat D\otimes \hat D ~~~\PPi~ \CF \right]
    &= \hat D\otimes \left(
      \left[\hat D ~~ \PPi\right] \CF + \PPi \left[\hat D ~~ \CF\right]\right)\label{eq:diagProd1}\\[4pt]
      &=\left[\hat D \otimes \hat D ~~ \PPi\right]\CF + \left[\I\otimes\hat
        D ~~\PPi \right]\cdot \left[\hat D\otimes\I ~~\CF \right]\nonumber\\&
     \quad +\left[\hat 
        D\otimes\I ~~\PPi \right]\cdot \left[\I\otimes \hat D ~~\CF \right] +
      \PPi \left[\hat D\otimes \hat D ~~\CF\right]\label{eq:diagProd2}\\[4pt]
      &=\gtD{\psdot(\xo,\yt) \psdot(\xt,\yt)}
      +\gtD{\psdot(\xo,\yo)\psdot(\xt,\yt)} \nonumber\\&
     \quad  +\gtD{\psdot(\xo,\yt)\psdot(\xt,\yo)}
      +\gtD{\psdot(\xo,\yo)\psdot(\xt,\yo)}\label{eq:diagProd3}\\[4pt]
      &=
{{\raisebox{-13pt}{
\renewcommand{\ll}{-2.3}\newcommand{\yo}{0.2}\newcommand{\rr}{2.3}   \newcommand{\yt}{0.5}\newcommand{\dd}{-0.2}\newcommand{\xo}{-1.3}\newcommand{\uu}{0.9} \newcommand{\xt}{.3}\newcommand{\ro}{1.2}\begin{pspicture}(\ll,\dd)(\rr,\uu)
          \psline[linecolor=blue](\ll,\yo)(\ro,\yo)
          \psline[linecolor=red](\ll,\yt)(\ro,\yt)
          \psline(\xo,\dd)(\xo,\uu)
          \psline(\xt,\dd)(\xt,\uu)
          \rput(1.7,\yt){\(_{\PPi}\)}
          \rput(1.7,\yo){\(_{\CF}\)}
          \rput(\xo,0.35){\RpR{red}{blue}}
          \rput(\xt,0.35){\RpR{red}{blue}}
        \end{pspicture}}}.~~}\label{eq:diagProd4}
    \end{align}
The equality between \eqref{eq:diagProd1} and \eqref{eq:diagProd2} is
just the Leibniz rule, while \eqref{eq:diagProd3} defines the graphical
representation of each term of \eqref{eq:diagProd2}. Each black dot
stands for a co-derivative, acting on what lies on its right on the
same horizontal line (horizontal lines actually hide auxiliary spaces
whose characters contribute to \(\PPi(g)\) [resp \(\CF\)] ). The vertical
lines correspond to the 
quantum space on which the whole operator acts, and the crossings
without dots stand for \(\I\). When operators are multiplied in the
quantum spaces, they are represented one above another. For instance 
\(\left[\I\otimes\hat D ~~\PPi \right]\cdot \left[\hat D\otimes\I ~~\CF
\right] = \gtD{\psdot(\xo,\yo)\psdot(\xt,\yt)}
\).

The last expression \eqref{eq:diagProd4} gives a shorter
representation of \eqref{eq:diagProd2}, where an implicit
multiplication is taken between the blocks \({\raisebox{-7pt}{\begin{pspicture}(0,-0.2)(0.0,.4)
  \RpR{red}{blue}
\end{pspicture}
}}\).
The generalization to an arbitrary number of spins \(N\) is straightforward
- one has \(N\) such blocks instead of two.

 One can also
see that due to the relation \(\hD \det(g)=\I \det(g)\) and to the Leibniz
rule, we have very generally 
\[\left[\bigotimes_i (u_1+2 \hD) ~~\PPi
\det(g)^a\right]=\det(g)^a  \left[\bigotimes_i (u_1+2 a+2 \hD)~~ \PPi\right] . 
\]

  \subsection{Commutativity of all  T- and Q-operators}
  \label{sec:commutation-relation}

Everywhere through this paper, the quantities of interest are of the
form\footnote{Most often, the operators of interest were actually of
  the form \(\left[\bigotimes_{i=1}^{N} (u_i + 2 ~\hat D
    )~~\PPi(g)\right]\), which is equivalent, after the rescaling
  \(u_i\to 2~ u_i\) and \(\Pi\to \Pi/2^N\), to \(\left[\bigotimes_{i=1}^{N} (u_i + \hat D
    )~~\PPi(g)\right]\).}
\(\left[\bigotimes_{i=1}^{N} (u_i + \hat D  )~~\PPi(g)\right]\), where \(\PPi(g)\)
is a class-invariant function of \(g\) (a symmetric function of
  its eigenvalues). 

In \cite{Kazakov:2007na}, the particular case when \(\PPi\) is the
character of a representation \(\{\l\}\) was studied, and in particular
it was proven that the T-operators commute
for different representations \(\{\l\}\) and \(\{\mu \}\), yielding the relation
%
\begin{align}
\left\llbracket\left[\bigotimes_{i=1}^{N} (u_i + \hat D  )~~\chi_{\{\l\}}(g)\right],\left[\bigotimes_{i=1}^{N} (u_i+v + \hat D
  )~~\chi_{\{\mu\}}(g)\right]\right\rrbracket&=0\label{eq:comm0},
\end{align}
where \(\left\llbracket A,B\right\rrbracket  \equiv
A\cdot B-B\cdot A\),  and 
\(u_{j},v \in {\mathbb C}\). 
Then, by writing an arbitrary class function as a linear combination
of characters, one gets
\begin{align}
\left\llbracket\left[\bigotimes_{i=1}^{N} (u_i + \hat D  )~~\PPi\right],\left[\bigotimes_{i=1}^{N} (u_i+v + \hat D
  )~~\PPi'\right]\right\rrbracket&=0\label{eq:comm},
\end{align}
which holds when \(\PPi, \PPi'\) are two arbitrary class functions of \(g\).

\section{\texorpdfstring{Degree in $u$ of the polynomials  $\Tb_{I}(u)$}{Degree in u of the polynomials  T(u)}}
\label{sec:redU}

We  already claimed that the l.h.s. of
  \eqref{eq:defQN1f} only has simple poles. Now that we have described
  diagrammatic rules for the action of co-derivative on \(w(z)\)
  functions, we can make this statement more explicit by writing the
  matrix coefficients of the Q-operator:

These matrix elements are given by a  generalization of the
formula (4.16) [(35) in the arXiv version] of \cite{Kazakov:2007na},
which reads (see also \eqref{eq:diag3})\footnote{Here, the function
  \(\theta(n)\) is equal to \(1\) 
  [resp 0] if \(n\ge 0\) [resp \(n\le -1\)]. 
%
}
\begin{multline}
  \left[\bigotimes_i (2+u_i+  
    2 \hD ) ~ w(t) \right]_{l_1,\,l_2,\,\,\cdots\, ,l_N}^{k_1,k_2,\cdots ,k_N}= 
    \\
 =
\sum_{\sigma\in \mathcal{S}_N}\prod_{i=1}^N \left(u_i\delta_{i,\sigma (i)}+\frac{2
    (g~t)^{\theta(\sigma(i)-i-1)}}{1-g~t}\right)^{k_i}_{l_{\sigma(i)}} w(t),
\label{eq:diagQ}
\end{multline}
where the sum is taken over the permutation group \(\mathcal{S}_{N}\)
of order \(N\). 
At this point, we can already notice that the only non-zero terms in
the sum are the permutations such that \(\forall i,
k_i=l_{\sigma(i)}\),
 since we are working in the diagonal basis. 
The same property holds for T-operators, and it
implies for instance that the number of spins pointing in each direction
\(|e_k\rangle\) is a conserved quantity.

After multiplication by 
\(\left(1-g
  t\right)^{\bigotimes N}\), the left-hand-side of \eqref{eq:defQN1f}
has only simple poles from \(\left.w(z)\right|_{z\to 1/x_k}\) :
\begin{multline}
\lefteqn{\left(\left(1-g t\right)^{\bigotimes N}\cdot \left[\bigotimes_i (u_i+2\hat 
    D+2)  ~w(t) \right]\right)_{l_1,\,l_2,\,\cdots\, l_N}^{k_1,k_2,\cdots
  k_N}=}
\\
=
\sum_{\sigma\in \mathcal{S}_N}\prod_{i=1}^N
\delta^{k_i}_{l_{\sigma(i)}}
 \left(u_i \delta_{i,\sigma(i)}~(1-x_{k_i}~t)+2
    (x_{k_i}~t)^{\theta(\sigma(i)-i-1)}\right) w(t).
\label{eq:QSum}
\end{multline}

In the definition of \(\Qb_{\overline \jmath}(u)\), we take the limit
\(t\to 1/x_j\), and we see that 
\(\Qb_{\overline \jmath}(u)|e_{k_1,k_2,\cdots k_N}\rangle\) 
is independent of all \(u_n\)
such that \(k_n=j\).
As a consequence, the degree in \(u\) of each eigenvalue of
\(\Qb_{\overline  \jmath}(u)\) is equal to the number of spins
pointing in the
directions \(|e_k\rangle_{k\neq j}\) in the corresponding eigenstate.

Moreover, the generalization to an arbitrary T or Q-operator is very
simple:\\ 
{ 
\(\Tb_{I}^{\{ \lambda \}}(u)|e_{l_1,l_2,\cdots , l_N}\rangle\) 
}
is independent of \(u_i\) for all \(i\) such that
\({l_i}\in \bI\), and the degree in \(u\) of each eigenvalue of 
\(\Tb_{I}^{\{ \lambda \}}(u)\) 
 is equal to the number of spins pointing in the
directions \(|e_j\rangle_{j\in I}\) in the corresponding eigenstate.

\section{Generating series of symmetric T-operators}
\label{Appx:TQ}

The TQ-relation \eqref{eq:TQg} can be re-written as an expression for
the generating series\footnote{From
  \(\chi_{s}(g_{\emptyset})=\delta_{s,0}\), we can 
  see that \(\mathfrak W_\emptyset(u,z)=\Qb_{\emptyset} (u)\), which we
  will use to get \eqref{eq:series}.} \(\mathfrak W_I(u,z)\equiv 
\sum _{s=0}^{\infty}z^s \Tb_{I}^s(u)\) of symmetric T-operators:\cite{Krichever:1996qd,Kazakov:2007fy,Zabrodin:2007rq,Tsuboi:1997iq}
\begin{align}
  \frac{\Qb_{I,j} (u)}{\Qb_{I} (u)}  \mathfrak W_I(u,z) &=
  \left(1 - x_{j}
    \frac{\Qb_{I} (u-2)}{\Qb_{I} (u)} z e^{2 \partial_u} \right) \mathfrak W_{I,j}(u,z), 
\label{eq:Wrec}
\end{align}
where we used a shift operator \(e^{2\partial_u}f(u) =f(u+2)\) for any function \(f(u)\). 
Let us then fix a chain of subsets of the full set: 
\(\F=I_{K} \supset I_{K-1} \supset \cdots \supset I_{0}=\emptyset \), 
where 
\(I_{k}=\{j_{1},j_{2},\cdots,j_{k}\}\), \(I_{k}\setminus
I_{k-1}=\{j_{k}\}\), 
\(k=1,2,\dots, K\). 
Then from \eqref{eq:Wrec}, we immediately get that
{ 
\begin{align}
\label{eq:series}
  \mathfrak W_{I_k}(u,z)&=\mathcal{O}_k
 \cdot
\mathcal{O}_{k-1} 
\cdots 
\mathcal{O}_{1} \cdot 
  \Qb_{\emptyset}(u), \\[6pt]
\textrm{where } 
\mathcal{O}_k&=\left(1 - x_{j_k}
    \frac{\Qb_{I_{k-1}} (u-2)}{\Qb_{I_{k-1}} (u)} z e^{2 \partial_u}
  \right)^{-1} \frac{\Qb_{I_{k}}(u)}{\Qb_{I_{k-1}}(u)}\nonumber
\\&=
  \sum_{n=0}^{\infty} \left( x_{j_k}\frac{\Qb_{I_{k-1}}
      (u-2)}{\Qb_{I_{k-1}} (u)} z e^{2 \partial_u}\right)^n \frac{\Qb_{I_{k}}(u)}{\Qb_{I_{k-1}}(u)},
      \nonumber 
\end{align}
}
which expresses the T-operators for symmetric tensor representations in terms
of the Q-operators \(\Qb_{I_{n}}\).
In particular, taking the coefficient of \(z\) in \eqref{eq:series}, 
one gets the usual relation for the fundamental
representation \(s=1\):
\begin{align}
\label{rec-trad}
\Tb^{1}_{I_{K}}(u)=
\Qb_{I_{K}}(u)
\sum_{m=1}^{K} 
x_{ { j_{m}} }
\frac{\Qb_{I_{m}}(u+2)}{\Qb_{I_{m}}(u)}\frac{\Qb_{I_{m-1}}(u-2)}{\Qb_{I_{m-1}}(u)}.
\end{align}

Note that this T-operator \eqref{rec-trad}  has the same form 
as  the T-function in 
\cite{Kulish:1983rd} 
obtained by the nested Bethe ansatz, if 
the Q-operators are replaced by their eigenvalues (Q-functions). 

The same generating series can also be written in the \(gl(K|M)\)
case 
for a fixed chain of subsets of the full set: 
\(\F=I_{K+M} \supset I_{K+M-1} \supset \cdots \supset I_{0}=\emptyset \), 
where 
\(I_{k}=\{j_{1},j_{2},\cdots,j_{k}\}\), \(I_{k}\setminus
I_{k-1}=\{ j_{k}\}\), 
\(k=1,2,\dots, K+M\): 
\eqref{eq:Wrec} still holds for \(j\leq K\), 
 while 
 for \(\hat l\ge
K+1\) \eqref{eq:TQf} gives :
\begin{align}
  \frac{\Qb_{I} (u)}{\Qb_{I,\hat l} (u)}  \mathfrak W_{I,\hat l}(u,z) &=
  \left(1 - y_{l}
    \frac{\Qb_{I,\hat l} (u-2)}{\Qb_{I,\hat l} (u)} z e^{2 \partial_u} \right) \mathfrak W_{I}(u,z),
\label{eq:Wrecf}
\end{align}
so that \eqref{eq:series} becomes for \(gl(K|M)\) :

\begin{align}
\label{eq:seriesS}
  \mathfrak W_{I_k}(u,z)&=\mathcal{O}_{k}\cdot
  \mathcal{O}_{{k-1}}\cdots
  \mathcal{O}_{1} \cdot \Qb_{\emptyset}(u), \\[6pt]
\textrm{where } 
\mathcal{\mathcal{O}}_{{ {k} }}&=\left(1 - x_{j_k}
    \frac{\Qb_{I_{k-1}} (u-2)}{\Qb_{I_{k-1}} (u)} z e^{2 \partial_u}
  \right)^{-1} \frac{\Qb_{I_{k}}(u)}{\Qb_{I_{k-1}}(u)} &&\mathrm{if~}
  j_k\leq K  ,\nonumber \\[6pt]
\mathcal{\mathcal{O}}_{ { {k}} }& =\frac{\Qb_{I_{k}}(u)}{\Qb_{I_{k-1}}(u)}
\left(1 - y_{l} 
    \frac{\Qb_{I_{k}} (u-2)}{\Qb_{I_{k}} (u)} z e^{2 \partial_u}
  \right) &&\mathrm{if~}
  j_k=\hat l > K
  \nonumber 
\end{align}
which gives for instance the following generalization of
\eqref{rec-trad} :
\begin{align}
\Tb^{1}_{I_{K+M}}(u)&=
\Qb_{\overline \emptyset}(u)
\sum_{k=1}^{K+M}(-1)^{p_{i_{k}}}\xi_{i_{k}} 
\frac{\Qb_{I_{k-1}}(u-2(-1)^{p_{i_{k}}})\Qb_{I_{k}}(u+2(-1)^{p_{i_{k}}})}{\Qb_{I_{k-1}}(u)\Qb_{I_{k}}(u)} . 
 \label{rec-tradsuper}
\end{align}
Note that 
the eigenvalue of \eqref{rec-tradsuper} coincides with a traditional form of the T-function 
from the Bethe ansatz  \cite{Sc83}. 
There are \((K+M) !\) ways to chose the chain \(\{I_{k}\}_{k=0}^{K+M}\), 
but \eqref{rec-tradsuper} does not 
depend on this choice. This is the 
(super) Weyl group symmetry of the T-operators.

\section{Bethe equations at the level of eigenvalues}
\label{sec:super-case-bethe}

As explained in section \ref{subsec:BAE}, the operator equation
\eqref{eq:OpBAE} can be written as a Bethe equation \eqref{BAE-eigen} on the
roots of 
the eigenvalues  \(\Qs_I(u)\) of the polynomial 
operators  \(\Qb_I(u)\)
for any eigenstate.

The same can be easily done in the supersymmetric case so that
\eqref{eq:OpBAE}, \eqref{opBAEff}, \eqref{eq:OBbf} and \eqref{eq:OBfb}
imply that for any eigenstate, the eigenvalues of the Q-operators satisfy
the conditions\footnote{Like in the section \ref{sec:super}, we use
  here the convention 
 \( i,j  \in \{1,2,\dots, K \}\), and denote by \(\hat l=l+K\) and
  \(\hat m=m+K\) for \( l,k \in  \{1,2,\dots, M\}\) 
 for some indices of fermionic grading.} :
\begin{align}
 -1&=\frac{x_{i}}{x_{j}} 
 \frac{\Qs_{I}(u^{(I,i)}_{k}-2)\Qs_{I,i}(u^{(I,i)}_{k}+2)\Qs_{I,i,j}(u^{(I,i)}_{k})}
{\Qs_{I}(u^{(I,i)}_{k})\Qs_{I,i}(u^{(I,i)}_{k}-2)\Qs_{I,i,j}(u^{(I,i)}_{k}+2)}
&
\; \; 
\text{for} & \; \; 1\le k \le K_{I,i}\;, 
 \label{BAEbb}\\[6pt]
 \label{BAEff}
  -1&=\frac{y_{l}}{y_{m}}
 \frac{\Qs_{I}(u_{k}^{(I,\hat{l})}+2) \Qs_{I,\hat{l}}(u_{k}^{(I,\hat{l})}-2) \Qs_{I,\hat{l},\hat{m}}(u_{k}^{(I,\hat{l})})}
 {\Qs_{I}(u_{k}^{(I,\hat{l})}) \Qs_{I,\hat{l}}(u_{k}^{(I,\hat{l})}+2) \Qs_{I,\hat{l},\hat{m}}(u_{k}^{(I,\hat{l})}-2)} 
&
 \; \;
 \text{for} & \; \; 
1\le k \le K_{I,\hat{l}} \; , \\[6pt]
1&=\frac{x_{i}}{y_{l}}
 \frac{\Qs_{I}(u_{k}^{(I,i)}-2) \Qs_{I,i,\hat{l}}(u_{k}^{(I,i)})}
{\Qs_{I}(u_{k}^{(I,i)}) \Qs_{I,i,\hat{l}}(u_{k}^{(I,i)}-2)}
&
 \; \;
 \text{for}& \; \; 
1 \le k \le  K_{I,i} \; , 
 \label{BAEbf}\\[6pt]
1&=\frac{y_{l}}{x_{i}}
 \frac{\Qs_{I}(u_{k}^{(I,\hat{l})}+2) \Qs_{I,i,\hat{l}}(u_{k}^{(I,\hat{l})})}
{\Qs_{I}(u_{k}^{(I,\hat{l})}) \Qs_{I,i,\hat{l}}(u_{k}^{(I,\hat{l})}+2)}
&
 \; \;
 \text{for}& \; \;
1 \le k \le K_{I,\hat{l}} \; , 
 \label{BAEfb}
\end{align}
where  
\((u_1^{(I)},\cdots,u_{K_I}^{(I)})\)
 are the roots of the polynomial \(\Qs_I(u)\).

These equations (\ref{BAEbb}-\ref{BAEfb}) are 
  equivalent to the traditional form of the Bethe equations 
\cite{BAE-ref}, respectively, [cf.\ eqs.\ (68), (69), (71) and (70) in \cite{Zabrodin:2007rq}.]
 

\section{Proof of the master identity \protect\eqref{eq:TQNu}} 
\label{sec:proof-mast-ident}

In this appendix, we will give a proof of the master identity \eqref{eq:TQNu} 
based on the so-called Bazhanov-Reshetikhin formula\footnote{We thank
  Anton Zabrodin who proposed us the idea to use the
  Bazhanov-Reshetikhin formula for the proof of our master identity.}.

Consider the main object of our Master identity the generating   operator of the transfer matrices  
\begin{align}
\W_{I}(u) = \bigotimes_{i}(u_{i}+ 2 \hat{D}) \prod_{k \in I}w(z_{k}), 
\end{align}
where \( z_{k} \in {\mathbb C}\) and \(I\) is any subset of \({\mathbb Z}_{>0}\). 
We assume \(z_{i} \ne z_{j}\)  for any \(i,j \in {\mathbb Z}_{>0}\) such that \(i\ne j\).
Then the master identity \eqref{eq:TQNu} can be rewritten in a form of
a Pl\"ucker identity (or Jacobi identity):
\begin{align}\label{eq:PlW}
(z_{i}-z_{j}) \W_{I,i,j}(u+2) \W_{I}(u)
=z_{i}\W_{I,j}(u)\W_{I,i}(u+2)
-z_{j}\W_{I,j}(u+2)\W_{I,i}(u),
\end{align}
where \(i,j \in {\mathbb Z}_{>0}\), \(i,j \notin I\), \(i \ne j\).
It can be solved  recursively in the same way as the QQ-relations   \eqref{eq:WronskQ}) giving 
\begin{align}
 \W_{I}(u) &= \frac{\det\left( z_{j}^{n-k} \W_{j}(u-2k+2) \right)_{1\le j,k \le n} }
{\det\left( z_{j}^{n-k} \right)_{1\le j,k \le n}
\prod_{k=1}^{n-1} \phi(u-2k) }.
  \label{master-det}
\end{align}
Here \(\phi(u)\equiv \W_{\emptyset}(u)=
\prod_{j=1}^{N}u_{j}=\Qb_{\overline \emptyset}(u)\), and
we consider (without losing generality) the case \(I=\{1,2,\dots,n\}\) 
for any finite \(n \in {\mathbb Z}_{> 0}\).

Hence, to prove the master identity \eqref{eq:TQNu} all we need is to prove this determinant formula \eqref{master-det}.   
 Let us expand\footnote{{ Eq.\ \eqref{expan-lhs} is an operator analogue of 
a generating function of the T-functions for any Young diagrams in} eq.\ (2.44) in
   \cite{Tsuboi:2009ud}.} the  
 left-hand side of \eqref{master-det} multiplied by a factor with respect to \(\{z_{k}\}\): 
\begin{multline}
\W_{I}(u) \det_{1\le j,k \le n} \left( z_{j}^{n-k} \right) 
 \prod_{j=1}^{n} z^{j-n}_{j}
=
\det_{1\le j,k \le n}\left( z_{j}^{j-k} \right) \W_{I}(u) 
 \\
= \bigotimes_{i}(u_{i}+2\hat{D}) 
\sum_{\sigma \in S_{n}} {\mathrm {sgn}} (\sigma)
\sum_{m_{1}=0}^{\infty} \sum_{m_{2}=0}^{\infty} \cdots \sum_{m_{n}=0}^{\infty} 
\prod_{k=1}^{n} \chi_{m_{k}}(g) 
z_{k}^{m_{k} -\sigma(k)+k}, 
 \label{expan-lhs} 
\end{multline}
where the sum\footnote{The sum over \(\{m_{k}\}_{k\in I}\) can be taken over any integers since 
\(\chi_{m}(g)=0\) if \(m<0\).}
 is taken over the permutation group \(S_{n}\) on the set \(\{1,2,\dots, n\}\) and \({\mathrm {sgn}} (\sigma)\) 
is a signature of the permutation \(\sigma\). 
The coefficient of \( \prod_{k=1}^{n} z_{k}^{\lambda_{k}}\) for any set of integers \(\{\lambda_{k}\}_{k \in I}\) 
 in \eqref{expan-lhs}, due to the Jacobi-Trudi formula
 \eqref{CHADET}, is nothing but the transfer matrix
 \(T_\lambda\) from \eqref{DIFFT} 
\begin{multline}
\bigotimes_{i}(u_{i}+2\hat{D}) 
\sum_{\sigma \in S_{n}}  {\mathrm {sgn}} (\sigma) 
\prod_{k=1}^{n} \chi_{\lambda_{k} +\sigma(k)-k}(g) 
= \\
=
\bigotimes_{i}(u_{i}+2\hat{D}) 
\det_{1\le j,k \le n } \left( \chi_{\lambda_{ j } +k-j }(g) \right).
\end{multline}
 Now let us expand the right hand side of \eqref{master-det} (times
 the same  \(\{z_{k}\}\)-dependent factor): 
\begin{multline}
\det_{1\le j,k \le n}\left( z_{j}^{n-k} \W_{j}(u-2k+2) \right) 
  \prod_{j=1}^{n} z^{j-n}_{j}
  \\
= 
\sum_{\sigma \in S_{n}} {\mathrm {sgn}} (\sigma)
\sum_{m_{1}=0}^{\infty} \sum_{m_{2}=0}^{\infty} \cdots \sum_{m_{n}=0}^{\infty} 
\prod_{k=1}^{n} \Tb^{m_{k}}(u-2 \sigma(k)+2) 
z_{k}^{m_{k} -\sigma(k)+k}.
 \label{expan-rhs} 
\end{multline}
The coefficient of \( \prod_{k=1}^{n} z_{k}^{\lambda_{k}}\) in \eqref{expan-rhs} is 
\begin{multline}
\sum_{\sigma \in S_{n}} {\mathrm {sgn}} (\sigma) 
\prod_{k=1}^{n} \Tb^{\lambda_{k} +\sigma(k)-k}(u-2 \sigma(k)+2) 
= \det_{1\le j,k \le n}\left( \Tb^{\lambda_{j} +k-j}(u-2k+2) \right) .
\end{multline}
Therefore, the proof of \eqref{master-det} is reduced to the following identity 
\begin{align} 
\bigotimes_{i}(u_{i}+2\hat{D}) 
\det_{1\le j,k \le n } \left( \chi_{\lambda_{ j } +k-j }(g) \right)
=
\frac{
\det_{1\le j,k \le n}\left( \Tb^{\lambda_{j} +k-j}(u-2k+2) \right)
}
{\prod_{k=1}^{n-1}\phi (u-2k)},
\label{BR-proof}
\end{align}
where the l.h.s. is precisely the T-operator \(\Tb^{ \{ \lambda \}}(u)\).   We recognize here the Bazhanov-Reshetikhin formula proven in \cite{Kazakov:2007na}. This proves the formulas \eqref{eq:PlW}-\eqref{master-det}, and hence the master identity \eqref{eq:TQNu}.\footnote{
Note that both sides of  \eqref{BR-proof} are antisymmetric w.r.t. the set  
\((\lambda_{1}-1, \lambda_{2}-2, \dots, \lambda_{n}-n)\). Hence, when
the determinant is non-zero, we can always relabel \(\l_k\)'s in such a way that  the highest weight components satisfy the usual inequalities:    \(\lambda_{1} \ge \lambda_{2} \ge \dots \ge \lambda_{n} \ge 0 \). 
 }

\section{Co-derivative and ``removal'' of eigenvalues}
\label{sec:Cre}
The co-derivative of \(\chi_{s}(g_I)\) a priori does not have such a
simple expression (as \eqref{eq:diag2} in terms of
diagrams) as the co-derivative of 
\(\chi_{s}(g)\). We will see in this subsection how to define the
action of co-derivative on \(\chi_{s}(g_I)\), and then we will see how
to compute the corresponding T-operators. In particular we will prove
the relation (\ref{defTc}).

\paragraph{Definition}~

The action of co-derivatives on \(w_I(z)\) (introduced in
(\ref{eq:chiwI})), can be defined by
means of equation (\ref{Ddef}), provided we specify what \(x_j\) is
at the point \(e^{\phi\cdot e}g\). 
The most natural definition is based on the fact that
\(x_j\) is the \(j^\mathrm{th}\) eigenvalue of \(g\), or
in other words the \(j^\mathrm{th}\) root of its characteristic
polynomial. In this sense, \(x_{j}\) is a function of the group element 
\(g\): \(x_{j}=x_{j}(g)\). In particular, \(x_{j}(\Omega g \Omega^{-1})=x_{j}(g)\) 
for any similarity transformation. 

If \(g\) is a diagonal matrix, it is immediate to see that 
the contribution of the non-diagonal-elements of the matrix
 \(e^{\phi\cdot e}g\) to 
 the characteristic polynomial \(\det\left(\l \I-e^{\phi\cdot e}g
\right) \) 
is at least  
quadratic in \(\phi\). This means that at the point \(e^{\phi\cdot e}g\),
\(x_j\) is equal to \(\left(e^{\phi\cdot  e}g\right)^j_j\) to the first order in \(\phi\). As a consequence, we get 
\(\hD_{j_1}^{i_1} x_j=\hD_{j_1}^{i_1} g_j^j=\de^{i_1}_{j}\de^{j}_{j_1}
x_j\), so that \(\hD x_j=\proj _j x_j\), where the projector to the eigenspace 
 for the \(j\)-th eigenvalue \(x_{j}\) is \(P_{j}=e_{jj}\) in this case.

More generally, if \(g=\Omega^{-1} \tilde g \Omega\) where \(\tilde g\) is
diagonal and \(\Omega\) is an arbitrary similarity transformation,
then\footnote{In \eqref{eq:Dre1}, we mainly use the relation (2.6) [equation (12) in 
the arXiv version] of \cite{Kazakov:2007na}.} 
we obtain 
\begin{align}
 \begin{split}
    \hD ~ x_j&= \frac{\partial}{\partial\phi} 
  \left.x_j\left(e^{\phi\cdot e}\Omega^{-1} \tilde g \Omega\right)\right|_{\phi=0}
  =\frac{\partial}{\partial\phi} 
  \left.\left(\Omega e^{\phi\cdot e}\Omega^{-1} \tilde g
  \right)^j_j\right|_{\phi=0}
\\[5pt]
  &=\sum_{i_{1}, j_{1}}e_{i_{1} j_{1}} \Omega^{j}_{j_{1}} (\Omega^{-1})^{i_{1}}_{j} x_{j} 
  =\sum_{i_{1}, j_{1}}e_{i_{1} j_{1}} (\Omega^{-1}e_{jj}\Omega )^{i_{1}}_{j_{1}} x_{j} .
  \end{split}
\label{eq:Dre1}
\end{align}

This exactly means that for a non-diagonal matrix \(g\), \(\hD ~ x_j =
\proj _j x_j\), where the projector to the eigenspace for \(x_{j}\) has the form 
\(P_{j}=\Omega^{-1} e_{jj} \Omega\).

\paragraph{Computation of T-operators}~

The claim which was already given in \eqref{defTc} is that the
computation of T-operators is done by commuting a factor
\(\frac{1}{w_{\bI}(z)}\) to the left of the co-derivatives.

{We will show that in the definition \eqref{defTc0} (or
equivalently \eqref{defTs}),} the multiplication by
\(\mathbf{B}_\bI\) introduced in \eqref{eq:QNorm} allows to commute any function of \(x_j\) (where \(j\in
\bI\)) across the co-derivatives. {As a consequence, the right
  hand sides of \eqref{defTc0} and \eqref{defTc} are equal, giving}
\begin{align}
\label{eq:commRM}
  \sum_{s=0}^{\infty} z^s \Tb_I^s(u) =
 \left(\prod_{j\in \bI}(1-x_j z)\right) \lim_{\substack{t_j\to \frac 1 {x_j}\\j\in
    \overline I}}
\mathbf{B}_\bI \cdot \left[\bigotimes_{i=1}^N (u_i+2 \hat D+2 |{\overline I}|)\,\,\,
    w(z)
 \PPi_\bI\right],
\end{align}
where the r.h.s. can be easily computed by diagrammatic methods.

In the case of \(1\) spin, this is checked by computing\footnote{Here, \( \left\llbracket A,B\right\rrbracket \) denotes the commutator \(AB-BA\).}
\begin{align}
\lim_{t\to\frac 1 {x_j}}  \left(1-g t\right)
{
\left\llbracket 
(u +2 \hat
  D),x_{j} \I 
\right\rrbracket
}
&=
\left(1-\frac g {x_j}\right) \cdot \left(2 \hat D x_j \right)
=2 \left(1-\frac g {x_j}\right)
x_j \proj_{j} \; = \; 0 . 
\label{eq:1spCm1}
\end{align}
We see that the key point in this commutation is the
multiplication by
\(\lim_{t\to\frac 1 {x_j}} (1-g t)= (1-g/x_j)\), which cancels the terms in \(\hat D
~x_j=x_j \proj_{j}\)
due to the property \( (1-g/x_j)~ \proj_{j}=0\). 

At \(N=m+n\) spins (\(m \in \mathbb{Z}_{\ge 0}\), \(n \in \mathbb{Z}_{\ge 1}\)), the analogous relation is 
\begin{align}
  C_{m,n}&=0 , \qquad\qquad 
\text{where} \quad 
C_{m,n}\equiv \left(1-g
    /x_j\right)^{\otimes (m+n)} B_{m,n}, \\[6pt]
B_{m,n}&\equiv
\left(\bigotimes_{i=1}^{m}(u_i+2\hat D)\right)\otimes 
\left\llbracket(u_{m+1} +2 \hat
  D),x_{j} \I
\right\rrbracket
 \otimes
\left(\bigotimes_{i=m+2}^{m+n}(u_i+2\hat D)\right), \nonumber
\end{align}
and it is proven by the recurrence over \(m\).
For 
\(m=0\), this follows from \eqref{eq:1spCm1}.
Let's show how \(C_{m+1,n}\) cancels under the assumption that
\(C_{m,n}=0\) for all \(g \in GL(K)\) 
 and any \(u_{j} \in {\mathbb C}\), \(j=1,2,\dots, m+n\).
Then for  any 
 \(u_{0} \in {\mathbb C} \), one can calculate: 
 \begin{align}
  0=&\left((1-g/x_j)\otimes \I^{\otimes (m+n)}\right)\cdot\left(
    (u_{0}+2\hat D) \otimes 
    C_{m,n}\right) \nonumber \\[6pt]
=&C_{m+1,n}^{\prime }+2\left((1-g/x_j)\otimes \I^{\otimes (m+n)} \right)\cdot 
\left[
\hat D  \otimes (1-g /x_j)^{\otimes (m+n)}
\right]
\cdot 
  (\I \otimes B_{m,n}), \label{eq:NspCm2}
\end{align}
where\footnote{In \eqref{eq:NspCm2}, we used the Leibniz rule
 \(\hD\otimes
  \left( \left(1-g/x_j\right)^{\otimes m+n} \cdot B_{m,n}\right) =
  \left[\hD\otimes 
  \left(1-g/x_j\right)^{\otimes m+n}\right]\cdot \left(\I\otimes
    B_{m,n}\right) + 
\left(\I\otimes \left(1-g/x_j\right)^{\otimes m+n}\right) \cdot
  \left[\hD\otimes B_{m,n}\right]
\)
.} 
\begin{align}
 \begin{split}
C_{m+1,n}^{\prime} & \equiv \left(1-g /x_j\right)^{\otimes (m+n+1)} B_{m+1,n}^{\prime}, \\[6pt]
B_{m+1,n}^{\prime}&\equiv\left(\bigotimes_{i=0}^{m}(u_i+2\hat D)\right)\otimes 
\left\llbracket(u_{m+1} +2 \hat
  D),x_{j} \I
\right\rrbracket
 \otimes
\left(\bigotimes_{i=m+2}^{m+n}(u_i+2\hat D)\right), 
 \end{split}
\end{align}
and due to\footnote{Here, the projector \(\proj_j\) on the the
  \(j^{\mathrm{th}}\) should not be confused with the permutation
  operator \(\mathcal{P}\) between the quantum spaces.} \(\hD\otimes g/x_j=\mathcal{P}\cdot(1\otimes
g/x_j)-\proj _j \otimes g/x_j\), the second term in
\eqref{eq:NspCm2} 
(multiplied by \(1/2\)) can be expanded to get
\begin{multline}
  -\frac 1 2 C_{m+1,n}^{\prime } =
  \\
= \left((1-g/x_j)\otimes \I^{\otimes (m+n)} \right)\cdot
\left(\sum_{k=1}^{m+n} \mathcal{P}_{0k}\cdot\I\otimes \left(\bigotimes_{i=1}^{m+n}
  (1-\delta_i^k -g/x_j ) \right)\right)\cdot 
  (\I \otimes B_{m,n}) 
\\[4pt]
-
\left((1-g/x_j)\otimes \I^{\otimes (m+n)} \right)\cdot
\left(\sum_{k=1}^{m+n} \proj_j\otimes \left(\bigotimes_{i=1}^{m+n}
  (1-\delta_i^k -g/x_j) \right)\right)\cdot   (\I \otimes B_{m,n}) \;,\nonumber
\end{multline}
and the first term becomes 
\begin{align}
-
\sum_k \mathcal{P}_{0k}\cdot \left(\I^{\otimes k}\otimes
(g/x_k)\otimes \I^{\otimes (n+m-k)}\right)
\cdot 
\left(\I\otimes
(1-g/x_k)^{\otimes (m+n)}\right)
\cdot  
\left(\I\otimes B_{m,n}\right), 
 \nonumber 
\end{align} 
which\footnote{Here, we use the fact that \(\left(
    (1-\frac g {x_j})\otimes \I^{\otimes (m+n)} \right) \cdot
  \mathcal{P}_{0,k} = \mathcal{P}_{0,k}\cdot \I^{\otimes
    k}\otimes(1-g/x_j)\otimes \I^{\otimes n+m-k}
\).} is zero because it contains \(C_{m,n}\).
The second term is also zero because it contains \((1-g/x_{j}) \proj_j\).
This completes the proof of the fact that 
\(C_{m+1,n}^{\prime}=0\),
 from which 
\(C_{m+1,n}=0\) follows. 

 As a consequence, we can indeed commute the factors \(\frac 1
 {{w_{\bI}(z)}}\) to the left of all co-derivatives in \eqref{defTc0}
 and get the relation \eqref{defTc}.

\pagebreak


\begin{thebibliography}{99}
\bibitem{Kazakov:2007na}
  V.~Kazakov and P.~Vieira,
  ``From Characters to Quantum (Super)Spin Chains via Fusion,''
  JHEP {\bf 0810} (2008) 050
  [arXiv:0711.2470 [hep-th]].

\bibitem{Bazhanov:1989yk}
  V.~Bazhanov and N.~Reshetikhin,
  ``Restricted Solid On Solid Models Connected With Simply laced Algebras And
  Conformal Field Theory,''
  J.\ Phys.\ A  {\bf 23}, 1477 (1990).


\bibitem{Pearce:1991ty}
A.~Kl{\"u}mper and P.~A.~Pearce,
{\it ``Conformal weights of RSOS lattice models and their fusion hierarchies''}
Physica A\ {\bf 183} (1992) 304.

\bibitem{Kuniba:1993cn}
A.~Kuniba, T.~Nakanishi and J.~Suzuki,
{\it ``Functional relations in solvable lattice models. 1: Functional relations
and representation theory,''}
Int.\ J.\ Mod.\ Phys.\ A {\bf 9} (1994) 5215
[arXiv:hep-th/9309137].


\bibitem{Kuniba:1995vc}
  A.~Kuniba, S.~Nakamura and R.~Hirota,
  J.\ Phys.\ A  {\bf 29}, 1759 (1996)
  [arXiv:hep-th/9509039].

\bibitem{Bazhanov:1996dr}
  V.~V.~Bazhanov, S.~L.~Lukyanov and A.~B.~Zamolodchikov,
  \textit{``Integrable Structure of Conformal Field Theory II. Q-operator and DDV
  equation,''}
  Commun.\ Math.\ Phys.\  {\bf 190} (1997) 247
  [arXiv:hep-th/9604044].

\bibitem{Tsuboi:1997iq}
  Z.~Tsuboi,
  \textit{ ``Analytic Bethe ansatz and functional equations for Lie superalgebra
  \(sl(r+1|s+1)\),''}
  J.\ Phys.\ A  {\bf 30}, 7975 (1997)
[arXiv:0911.5386 [math-ph]].
  
\bibitem{Gromov:2008gj}
  N.~Gromov, V.~Kazakov and P.~Vieira,
  \textit{ ``Finite Volume Spectrum of 2D Field Theories from Hirota Dynamics,''}
  JHEP {\bf 0912}, 060 (2009)
  [arXiv:0812.5091 [hep-th]].
  
 
  \bibitem{Kazakov:2010kf}
  V.~Kazakov and S.~Leurent,
  \textit{``Finite Size Spectrum of SU(N) Principal Chiral Field from Discrete Hirota
  Dynamics,''}
  arXiv:1007.1770 [hep-th].
  
\bibitem{Gromov:2009tv}
  N.~Gromov, V.~Kazakov and P.~Vieira,
  \textit{ ``Exact Spectrum of Anomalous Dimensions of Planar N=4 Supersymmetric
  Yang-Mills Theory,''}
  Phys.\ Rev.\ Lett.\  {\bf 103} (2009) 131601 [arXiv:0901.3753 [hep-th]].

\bibitem{Bombardelli:2009ns}
  D.~Bombardelli, D.~Fioravanti and R.~Tateo,
  \textit{ ``Thermodynamic Bethe Ansatz for planar AdS/CFT: a proposal,''}
  J.\ Phys.\ A  {\bf 42}, 375401 (2009)
  [arXiv:0902.3930 [hep-th]].
 \(\bigstar\) 
  N.~Gromov, V.~Kazakov, A.~Kozak and P.~Vieira,
  \textit{ ``Integrability for the Full Spectrum of Planar AdS/CFT II,''}
 Lett.\ Math.\ Phys.\ {\bf 91}, 265 (2010)
  [arXiv:0902.4458 [hep-th]].
 \(\bigstar\)
  G.~Arutyunov and S.~Frolov,
  \textit{ ``Thermodynamic Bethe Ansatz for the \(AdS_5 \times S^5\) Mirror Model,''}
  JHEP {\bf 0905}, 068 (2009)
  [arXiv:0903.0141 [hep-th]].



\bibitem{Cherednik}
I.~Cherednik, {\it On special basis of irreducible representations of
degenerated affine Hecke algebras}, Funk. Analys. i ego Prilozh. {\bf 20:1}
(1986) 87-88 (in Russian) \(\bigstar\)
I.~Cherednik, {\it  Quantum groups as hidden symmetries of classical
representation theory }, Proceed. of 17th Int. Conf. on diff. geom. methods
in theoretical physics, World Scient. (1989), {\bf 47}.
\(\bigstar\)
I.~Cherednik, {\it  On irreducible representations of elliptic
quantum R-algebras}, Dokl. Akad. Nauk SSSR 291:1, 49-53 (1986)
Translation: M 34-1987, 446-450.
\(\bigstar\)
I.~Cherednik, {\it  An analogue of character formula for Hecke
algebras}, Funct. Anal. and Appl. 21:2, 94-95 (1987) (translation:
pgs 172-174).





  \bibitem{Bax72}
R.J.\ Baxter, 
Partition function of the eight-vertex lattice model, 
 Ann. Phys. 70 (1972) 193-228.

\bibitem{Bazhanov:1998dq}
  V.~V.~Bazhanov, S.~L.~Lukyanov and A.~B.~Zamolodchikov,
  \textit{ ``Integrable structure of conformal field theory. III: The Yang-Baxter
  relation,''}
  Commun.\ Math.\ Phys.\  {\bf 200}, 297 (1999)
  [arXiv:hep-th/9805008].




\bibitem{BaxQ-papers}
V.\ Pasquier, M.\ Gaudin, 
The periodic Toda chain and a matrix generalization of the 
Bessel function recursion relations, 
J. Phys. A: Math. Gen. 25 (1992) 5243-4252;
\(\bigstar\) 
%
K.\ Hikami, 
Baxter Equation for Quantum Discrete Boussinesq Equation, 
Nucl. Phys. B604 (2001) 580-602 [arXiv:nlin/0102021]; 
\(\bigstar\) 
%
K.\ Fabricius, B.M.\ McCoy,
New Developments in the Eight Vertex Model, 
J.Statist.Phys. 111 (2003) 323-337 [arXiv:cond-mat/0207177]; 
\(\bigstar\) 
%
V.B.\ Kuznetsov, V.V.\ Mangazeev, E.K.\ Sklyanin,
Q-operator and factorised separation chain for Jack polynomials, 
Indag. Math. 14 (2003) 451-482 [arXiv:math/0306242[math.CA]];
\(\bigstar\) 
%
  P.~P.~Kulish, A.~M.~Zeitlin,
  ``Superconformal field theory and SUSY N=1 KDV hierarchy II: The Q-operator,''
  Nucl.\ Phys.\  {\bf B709 } (2005)  578 
  [hep-th/0501019]; 
\(\bigstar\)
%
C.\ Korff, 
A Q-Operator Identity for the Correlation Functions of the Infinite XXZ Spin-Chain, 
J.Phys. A: Math. Gen. 38 (2005) 6641-6658 
[arXiv:hep-th/0503130];
 \(\bigstar\) 
%
A.G.\ Bytsko, J.\ Teschner, 
Quantization of models with non-compact quantum group symmetry. Modular XXZ magnet and lattice sinh-Gordon model, J.Phys.A 39 (2006) 12927-12981[arXiv:hep-th/0602093];
 \(\bigstar\)
%
H.\ Boos, M.\ Jimbo, T.\ Miwa, F.\ Smirnov, Y.\ Takeyama, 
 Hidden Grassmann structure in the XXZ model, 
Commun. Math. Phys. 272 (2007) 263-281 [arXiv:hep-th/0606280]; 
%
 \(\bigstar\) 
T.\ Kojima,
\textit{``The Baxter's Q-operator for the W-algebra \(W_N\),''}
J.Phys.A: Math. Theor. 41 (2008) 355206
[arXiv:0803.3505 [nlin.SI]].
%
 \(\bigstar\) 
S.E.\ Derkachov, A.N.\ Manashov, 
Factorization of R-matrix and Baxter Q-operators for generic \(sl(N)\) spin chains, 
J.Phys.A: Math. Theor. 42 (2009) 075204; [arXiv:0809.2050 [nlin.SI]].
%
 \(\bigstar\)  
H.\ Boos, F.\ G\"{o}hmann, A.\ Kl\"{u}mper, K.S.\ Nirov, A.V.\ Razumov, 
Exercises with the universal R-matrix 
J. Phys. A: Math. Theor. 43 (2010) 415208 [arXiv:1004.5342 [math-ph]]. 
%
 \(\bigstar\) 
  V.~V.~Bazhanov, T.~Lukowski, C.~Meneghelli and M.~Staudacher,
  {\it ``A Shortcut to the Q-Operator,''}
  J.\ Stat.\ Mech.\ {\bf 1011} (2010) P11002 
%
  [arXiv:1005.3261 [hep-th]].
 \(\bigstar\)
S.E.\ Derkachov, A.N.\ Manashov,
\textit{``Noncompact \(sl(N)\) spin chains: Alternating sum representation for finite dimensional transfer matrices,''}
arXiv:1008.4734 [nlin.SI].


\bibitem{Bazhanov:2001xm}
  V.~V.~Bazhanov, A.~N.~Hibberd and S.~M.~Khoroshkin,
  ``Integrable structure of \(W_{3}\) conformal field theory, quantum Boussinesq
  theory and boundary affine Toda theory,''
  Nucl.\ Phys.\  B {\bf 622} (2002) 475
  [arXiv:hep-th/0105177].


\bibitem{Belitsky:2006cp}
  A.~V.~Belitsky, S.~E.~Derkachov, G.~P.~Korchemsky and A.~N.~Manashov,
  \textit{ ``Baxter Q-operator for graded \(SL(2|1)\) spin chain,''}
  J.\ Stat.\ Mech.\  {\bf 0701}, P005 (2007)
  [arXiv:hep-th/0610332].


\bibitem{Bazhanov:2008yc}
  V.~V.~Bazhanov and Z.~Tsuboi,
  \textit{ ``Baxter's Q-operators for supersymmetric spin chains,''}
  Nucl.\ Phys.\  B {\bf 805}, 451 (2008)
  [arXiv:0805.4274 [hep-th]];
in section 2.4 of this paper, 
solutions of the Yang-Baxter relation (L-operators) for the Q-operators  
in the  \(U_{q}(\widehat{sl}(2|1))\) case  
were presented and the Q-operators were given as the (super)trace of these L-operators over some 
oscillator representations. 
 An idea of the derivation of these solutions of the Yang-Baxter relation for the Q-operators 
was presented at the
{\it ``Workshop and Summer School: 
From Statistical Mechanics to Conformal and Quantum Field Theory'', 
the university of Melbourne, January, 2007}
 and 
{\it {La 79eme Rencontre entre physiciens theoriciens et mathematiciens 
``Supersymmetry and Integrability'', IRMA Strasbourg, June, 2007  
.}
}


\bibitem{BFLMS10}
  V.~V.~Bazhanov, R.~Frassek, T.~Lukowski, C.~Meneghelli and M.~Staudacher, 
%
{\it ``Baxter Q-Operators and Representations of Yangians
,''}
 arXiv:1010.3699 [math-ph]. 
%
The results of this paper have been presented as a talk 
of M.Staudacher at a conference ``Integrability in Gauge and String Theory 2010''
(Nordita, Sweden, 28 June 2010).
  
\bibitem{Kazakov:2007fy}
  V.~Kazakov, A.~S.~Sorin and A.~Zabrodin,
  \textit{ ``Supersymmetric Bethe ansatz and Baxter equations from discrete Hirota
  dynamics,''}
  Nucl.\ Phys.\  B {\bf 790}, 345 (2008)
  [arXiv:hep-th/0703147].

\bibitem{Krichever:1996qd}
  I.~Krichever, O.~Lipan, P.~Wiegmann and A.~Zabrodin,
  \textit{``Quantum integrable models and discrete classical Hirota equations,''}
  Commun.\ Math.\ Phys.\  {\bf 188} (1997) 267
  [arXiv:hep-th/9604080].
  
\bibitem{Zabrodin:2007rq}
  A.~Zabrodin,
  ``Backlund transformations for difference Hirota equation and supersymmetric
  Bethe ansatz,''
  Theor. Math. Phys. {\bf 155}, 567 (2008) 
  [arXiv:0705.4006 [hep-th]].
  
\bibitem{Tsuboi:2009ud}
  Z.~Tsuboi,
  \textit{ ``Solutions of the T-system and Baxter equations for supersymmetric spin chains,''}
  Nucl.\ Phys.\  B {\bf 826} (2010) 399
  [arXiv:0906.2039 [math-ph]].


\bibitem{QQ-boson}
G.P.\ Pronko, Yu.G.\ Stroganov,   
\textit{ ``Families of solutions of the nested Bethe Ansatz for the \(A_2\) spin chain,''} 
J. Phys. A: Math. Gen. 33 (2000) 8267-8273 
[arXiv:hep-th/9902085]; 
%
%
\(\bigstar\)
P.\ Dorey, C.\ Dunning, D.\ Masoero, J.\ Suzuki, R.\ Tateo, 
\textit{ ``Pseudo-differential equations, and the Bethe Ansatz for the classical
Lie algebras,''}  
Nucl. Phys. B 772 (2007) 249-289 [arXiv:hep-th/0612298].

\bibitem{GS03}
F.\ G\"ohmann, A.\ Seel: 
\textit{ ``A note on the Bethe Ansatz solution of the supersymmetric t-J model,''} 
Czech.J.Phys. 53 (2003) 1041
[arXiv:cond-mat/0309138].

\bibitem{Gromov:2007ky}
  N.~Gromov and P.~Vieira,
  \textit{ ``Complete 1-loop test of AdS/CFT,''}
  JHEP {\bf 0804}, 046 (2008)
  [arXiv:0709.3487 [hep-th]].


\bibitem{Woynarovich83}
F.\ Woynarovich, 
\textit{ ``Low-energy excited states in a Hubbard chain with on-site attraction,''} 
J.Phys.C: Solid State Phys. 16 (1983) 6593; 
 \(\bigstar\)
%
P.A.\ Bares, I.M.P.\ Carmelo, J.\ Ferrer, P.\ Horsch,  
\textit{ ``Charge-spin recombination in the one-dimensional supersymmetric t-J model,''}  
  Phys. Rev. B46 (1992) 14624. 
%
\(\bigstar\)
Z.~Tsuboi,
\textit{ ``Analytic Bethe Ansatz and functional equations associated with
any simple root systems of the Lie superalgebra \(sl(r+1|s+1)\)''},
  Physica A {\bf 252}, 565 (1998)
[arXiv:0911.5387 [math-ph]].

\bibitem{BAE-ref}
C.K.\ Lai, 
``Lattice gas with nearest neighbor interaction in one dimension with arbitrary statistics,''
J.\ Math.\ Phys.\ {\bf 15} (1974) 1675. 
%
\(\bigstar\) 
  B.~Sutherland,
  ``Model for a multicomponent quantum system,''
  Phys.\ Rev.\  {\bf B12 } (1975)  3795.
  
\bibitem{Kulish:1983rd}
O.\ Babelon, H.J.\ de Vega, C-M.\ Viallet, 
  \textit{``Exact solution of the \(Z_{n+1}\times Z_{n+1}\) symmetric generalization of the
 XXZ model,''}  
Nucl. Phys. {\bf B200} (1982) 266. 
 \(\bigstar\) 
  P.~P.~Kulish, N.~Y.~.Reshetikhin,
  \textit{``Diagonalization Of \(Gl(n)\) Invariant Transfer Matrices and Quantum \(N\) Wave System (Lee Model),''}
  J.\ Phys.\ A {\bf A16 } (1983)  L591.

\bibitem{DM92}
T.\ Deguchi, P.P.\ Martin,
\textit{``An Algebraic Approach to Vertex Models and Transfer-Matrix Spectra, ''}
Int.\ J.\ Mod.\ Phys.\ {\bf A7}, Suppl. 1A (1992) 165.

\bibitem{Gromov:2010km}
  N.~Gromov, V.~Kazakov, S.~Leurent and Z.~Tsuboi,
  ``Wronskian Solution for AdS/CFT Y-system,'' 
  JHEP {\bf 1101} (2011) 155 
  [arXiv:1010.2720 [hep-th]].


\bibitem{Faddeev:1996iy}
  L.~D.~Faddeev,
  arXiv:hep-th/9605187.
\(\bigstar\)
Andreas Kl\"umper, {\it Integrability of quantum chains: theory and
applications to the spin-1/2 XXZ chain}, arXiv:cond-mat/0502431.




\bibitem{Gromov:2010vb}
  N.~Gromov, V.~Kazakov and Z.~Tsuboi,
\textit{  ``\(PSU(2,2|4)\) Character of Quasiclassical AdS/CFT,''
}JHEP{\bf 1007}(2010)097 [arXiv:1002.3981 [hep-th]].

\bibitem{Hegedus:2009ky}
  A.~Hegedus,
  \textit{ ``Discrete Hirota dynamics for AdS/CFT,''}
  Nucl.\ Phys.\  B {\bf 825}, 341 (2010)
  [arXiv:0906.2546 [hep-th]].

\bibitem{Bazhanov:1996aq}
  V.~V.~Bazhanov, S.~L.~Lukyanov and A.~B.~Zamolodchikov,
  ``Quantum field theories in finite volume: Excited state energies,''
  Nucl.\ Phys.\  B {\bf 489} (1997) 487
  [arXiv:hep-th/9607099].
  
\bibitem{Dorey:1996re}
  P.~Dorey and R.~Tateo,
  ``Excited states by analytic continuation of TBA equations,''
  Nucl.\ Phys.\  B {\bf 482}, 639 (1996)
  [arXiv:hep-th/9607167].

\bibitem{Gromov:2009zb}
  N.~Gromov, V.~Kazakov and P.~Vieira,
  \textit{ ``Exact AdS/CFT spectrum: Konishi dimension at any coupling,''}
Phys.\ Rev.\ Lett.\ {\bf 104}, 211601 (2010)
  [arXiv:0906.4240 [hep-th]].

\bibitem{Sc83}
C.L.\ Schultz, 
 \textit{``Eigenvectors of the multicomponent generalization of
the six-vertex model,''} 
Physica {\bf A122} (1983) 71.


\end{thebibliography}
\end{document}